\documentclass[preprint,aps,prd,showpacs,superscriptaddress,nofootinbib]{revtex4}
\usepackage{slashed}
\usepackage{amsfonts,graphicx,bm}
\usepackage{amsmath}
\usepackage{amssymb}
\usepackage{multirow}

\def\met{\displaystyle{\not}E_T}
\let\jnfont=\rm
\def\NPB#1,{{\jnfont Nucl.\ Phys.\ B }{\bf #1},}
\def\PLB#1,{{\jnfont Phys.\ Lett.\ B }{\bf #1},}
\def\EPJC#1,{{\jnfont Eur.\ Phys.\ Jour.\ C }{\bf #1},}
\def\PRD#1,{{\jnfont Phys.\ Rev.\ D }{\bf #1},}
\def\PRL#1,{{\jnfont Phys.\ Rev.\ Lett.\ }{\bf #1},}
\def\MPLA#1,{{\jnfont Mod.\ Phys.\ Lett.\ A }{\bf #1},}
\def\JPG#1,{{\jnfont J.\ Phys.\ G }{\bf #1},}
\def\CTP#1,{{\jnfont Commun.\ Theor.\ Phys.\ }{\bf #1},}

\begin{document}

\title{Search for $W^\prime$ signal via $tW^\prime$ associated production  at LHC}

\author{Xue Gong}
\affiliation{School of Physics, Shandong University, Jinan, Shandong 250100,  China}

\author{Hong-Lei Li}\email{sps$_$lihl@ujn.edu.cn}
\affiliation{School of Physics and Technology, University of Jinan, Jinan Shandong 250022,  China}
\affiliation{Department of Physics, University of the Chinese Academy of Sciences, Beijing 100049,  China}

\author{Cong-Feng Qiao}
\affiliation{Department of Physics, University of the Chinese Academy of Sciences, Beijing 100049,  China}

\author{Zong-Guo Si}
\affiliation{School of Physics, Shandong University, Jinan, Shandong 250100,  China}

\author{Zhong-Juan Yang}
\affiliation{School of Physics and Chemistry, Henan Polytechnic University, Jiaozuo Henan, 454003, China}

\begin{abstract}
A variety of new physics models predict the existence of extra charged gauge bosons ($W^\prime$). It is verified that $W^\prime$ and top quark associated production is a promising  process to search for $W^\prime$ signal at the LHC. We study the collider signatures of multi-jets$+$lepton$+\met$ by reconstructing the $tW'$ intermediate state  through the decay modes of $W^\prime \to tb$, $W^\prime \to q \bar q'$ and $W^\prime \to l \nu$ respectively. An angular distribution related to charged lepton and top quark moving direction is provided to distinguish alternative left-handed or right-handed chiral couplings of $W^\prime$ to quarks in the hadronic decay modes. The superiority on the search of $W^\prime$ is demonstrated in the leptonic decay channel for the left-handed type interaction, which is forbidden for  the right-handed type interaction in the most theories. To be more realistic, the relevant standard model backgrounds are simulated. We adopt various kinematic cuts to suppress the backgrounds and
collect the integral luminosity needed for the corresponding process detected at the LHC with $3\sigma$ sensitivity. We also provide a forward-backward asymmetry which is related to the chirality of $W^\prime$. The successive studies can shed light on the potential searching for  $W^\prime$ signal as well as distinguishing typical new physics models.
\end{abstract}
\maketitle
\section{Introduction}

Though the standard model (SM) has gained great success and SM-like Higgs boson is discovered at the LHC, it is still in progress to investigate the new phenomena induced by
many new physics models, such as extra dimensional models~\cite{Klein:1926tv,ArkaniHamed:1998rs,Randall:1999vf,ArkaniHamed:2001ca}, grand unified theories~\cite{ Pati:1973uk,Georgi:1974sy,Fritzsch:1974nn} and left-right symmetric models~\cite{Pati:1974yy,Mohapatra:1974hk,Mohapatra:1974gc,Senjanovic:1975rk,Mohapatra:1977mj}, etc. Among this kind of models, a new extra charged gauge boson $W^\prime$ is proposed. To discriminate different new physics models beyond the SM,  it is crucial  to search for the $W^\prime$ production signal and study its properties at the LHC.

Recently, the latest experimental results have explored the potential to observe the heavy gauge bosons at the LHC.  Searching for a $W^\prime$ boson with a signature of lepton and  missing transverse energy has been performed by ATLAS and CMS collaborations. The results show that no significant excess over the SM expectation has been observed through $W^\prime \to e\nu $ or $W^\prime \to \tau \nu$ decay~\cite{Aad:2012dm,Chatrchyan:2013lga}. On the other hand, the signature of $W^\prime \to  tb$ has also been investigated at the center of mass energy of 7 and 8 TeV, and the observed limits are displayed on the cross section as well as $W^\prime$ mass~\cite{Aad:2012ej,Chatrchyan:2012gqa,atlas8tev}.

In general, $W^\prime \to l \nu$ is the most prospective channel for its research at the LHC due to the unambiguous backgrounds, while $W^\prime \to tb$ decay channel also becomes important, especially within the models in which the couplings of $W^\prime$ to leptons are extremely suppressed.  The collider signature of a top-philic $W^\prime$, which couples only to the third generation quarks of the SM, produced in association with a top quark is investigated at the LHC~\cite{Berger:2011xk}. In addition, the investigation on the properties of $W^\prime$ also contributes to searching for its signal at the LHC, such as the chirality of $W^\prime$ coupling to SM particles. In reference~\cite{Gopalakrishna:2010xm}, $pp\to W^\prime \to tb$  process is analyzed and the authors propose that the angular distribution for the charged lepton decayed from the top quarks can be a feature of the chirality of $W^\prime$ with $W^\prime \to tb$ decay mode. The extensively studies on the chiral property are carried out in
the gauge interaction of $W^\prime$ decaying into $W$ and Higgs boson, which can be used to distinguish $W^\prime$ from the charged Higgs boson~\cite{Bao:2011nh,Bao:2011sy}.

In this paper we study the $W^\prime$ production in association with top quark at the LHC and the future hadron colliders. The dominant three decay modes of $W^\prime \to tb$, $W^\prime \to q \bar {q}'$ ($q$ stands for the light quark) and $W^\prime \to l \nu$ are investigated with the collider signature of $5{\rm jets}+l^{\pm}+\met$, $3{\rm jets}+l^{+}+\met$  and $3{\rm jets}+l^{-}+\met$. Searching for a massive $W^\prime$ signature with the electroweak coupling to fermions through the $W^\prime \to tb$ decay channel is hard due to the large backgrounds. The $W^\prime \to q \bar {q}'$ channel shows evidence in the search of $W^\prime$ at $\sqrt{s}=33$  TeV. Because the highly boosted charged lepton decayed from $W'$ differs from the backgrounds, the investigation of $W^\prime \to l \nu$ channel contributes to the search of $W'$ signal. Moreover, we provide the charged lepton angular distribution to distinguish the  chirality of $W^\prime$.

This paper is organized as follows. The couplings of $W^\prime$ to SM particles are discussed in Sec. II together with the mass constraints. In Sec. III, the numerical results with $W^\prime \to tb$, $W^\prime \to q \bar q'$ and $W^\prime \to l \nu$ decay modes are listed respectively and the corresponding SM backgrounds are simulated. Finally, we give a brief summary in Sec. IV.

\section{Theoretical Framework and Mass Constraints}
\label{secii}

The heavy charged gauge boson appears in various models with different couplings. We  extract the $W^\prime$ couplings to quark and lepton ($\psi$) from  the following general formula,
\begin{equation}
{\cal L} = \frac{ g_{L} }{\sqrt{2}}V_{L}^{\prime ij}\bar{\psi_u^i}\gamma_{\mu}
P_{L}\psi_d^j{{W'^{+}_{L}}^{\mu}}
+\frac{ g_{R}}{\sqrt{2}}V_{R}^{\prime ij}\bar{\psi_u^i}\gamma_{\mu}
P_{R}\psi_d^j{{W'^{+}_{R}}^{\mu}}
 + {\rm h.c.} \ ,
\label{wffcoupling}
\end{equation}
where $g_{L} (g_{R})$ is the coupling constant, $V_{L}^{\prime}$ ($V_{R}^{\prime}$) is a unitary matrix representing the fermion flavor mixing, and $P_{L,R}=(1\mp \gamma_5)/2$ is the left-, right-handed chiral projection operator. To give a simplified result, we set $g_L=g_2,~g_R=0$ with the pure left-handed gauge boson $W^\prime_L$, $g_L=0,~g_R=g_2$ with the pure right-handed gauge boson $W^\prime_R$, where $g_2$ is the electro-weak coupling constant in the SM.

The mass parameter is one of the most crucial factors in the search of $W^\prime$ as well as the interaction couplings.
From the proton anti-proton collisions, CDF and D$\emptyset$ collaborations obtain the lower bound for a SM-like $W^\prime$ at 0.8 and 1 TeV ~\cite{Aaltonen:2009qu,Abazov:2007ah},  respectively. The investigations on the K and B meson demonstrate that the $W^\prime$ lower mass limit is around 2.5 TeV ~\cite{Zhang:2007da}. The recent searches of $W^\prime$ have been performed by the ATLAS and CMS detectors at the LHC. Based on the results of $W^\prime \to l \nu$ decay channel, a $W^\prime$ with sequential SM couplings is excluded at the 95$\%$ credibility level for masses up to 2.55 and 2.90 TeV ~\cite{Aad:2012dm,Chatrchyan:2013lga}. As discussed in the references \cite{Mohapatra:1974hk,Mohapatra:1974gc,Senjanovic:1975rk,Mohapatra:1977mj}, one can imagine neutrino being Dirac particle. In this case, $W^{\prime}_R$ would have the usual leptonic decay channel, which implies a lower limit on its mass about 3 TeV. Independent of  the nature of neutrino mass, the left-right symmetric theory implies a theoretical lower limit on the $W^\prime_R$ mass of about 2.5 TeV \cite{Zhang:2007da,Beall:1981ze,Maiezza:2010ic}. Additionally, the right-handed $W^\prime$  with the mass below 1.13 and 1.85 TeV is excluded through the reconstructed $tb$ resonances ~\cite{Aad:2012ej,Chatrchyan:2012gqa,atlas8tev}. The search for the dijet mass spectrum, including quark-quark, quark-gluon, and gluon-gluon pairs, also presents mass bound $M_{W^{\prime}}>1.51$ TeV~\cite{Chatrchyan:2011ns}. In
particular, the CMS detector illustrates the mass of right-handed $W^\prime$ should be larger than 0.84 TeV through the $tW^{\prime}$ associated production ~\cite{Chatrchyan:2012su}.
\begin{figure}
\centering
\includegraphics[width=0.40\textwidth]{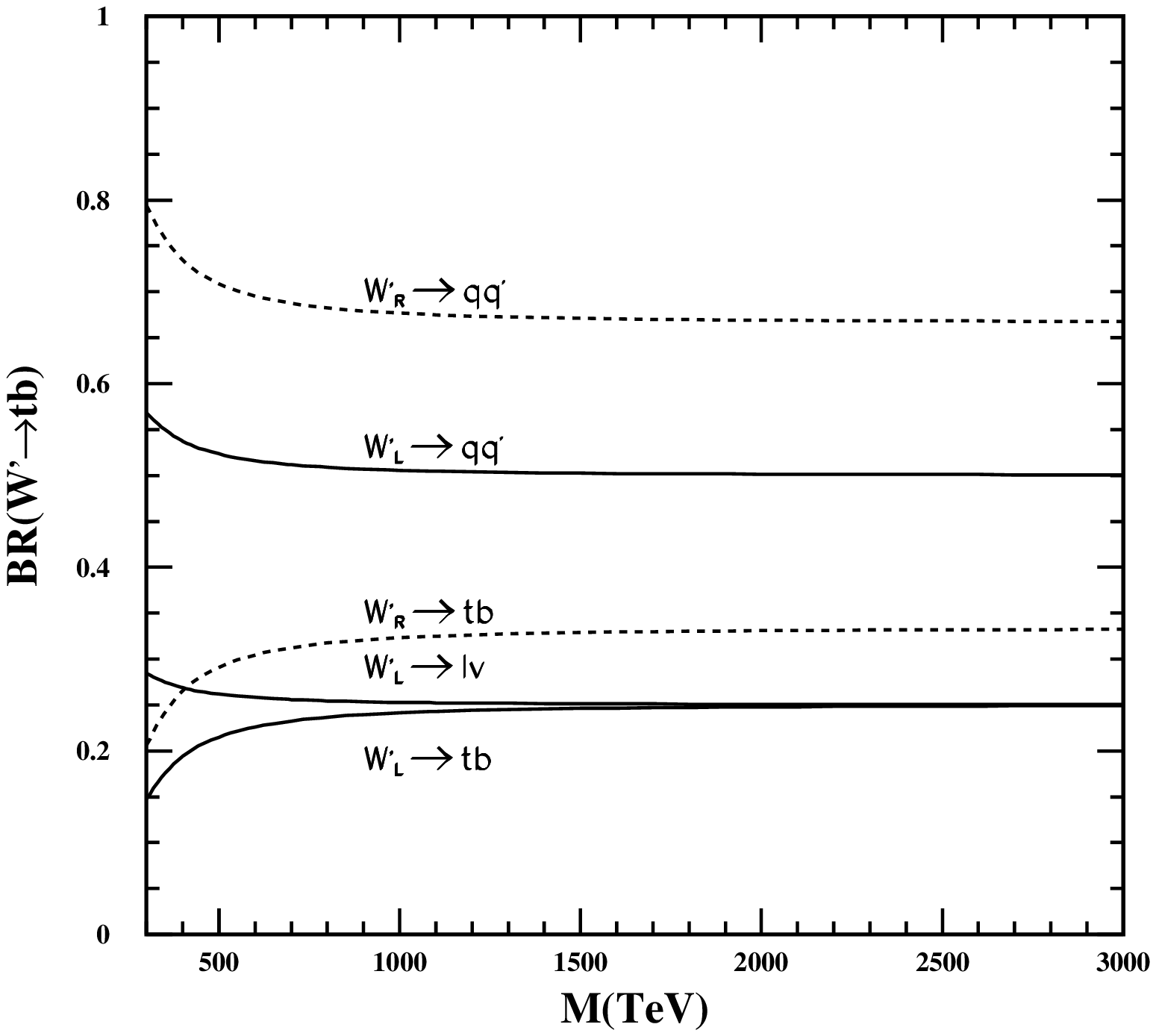}
\caption{The branching fractions of $W^\prime \to tb$,$W^\prime \to q\bar q^{\prime}$,
and $W^\prime \to lv$.}
\label{fig:ratio}
\end{figure}

The branching fractions of $W^\prime$ decay into fermions are shown in Fig.~\ref{fig:ratio}. Whatever the left-handed or right-handed $W^\prime$ decay into light quarks leads to more than a half fractions in the massive region. A disparate decay mode between the left-handed and right-handed $W^\prime$ is that the $W^\prime _R \to l \nu$ is forbidden due to only left-handed neutrino existed in the SM. We completely study the signatures of $W^\prime$ production in association with top quarks via the three decay modes, $W^\prime \to tb$, $W^\prime \to q\bar q'$ and $W^\prime \to l\nu$ in the following section.

\section{Numerical Results and discussion}
\label{seciii}

\begin{figure}
\centering \includegraphics[width=0.6\textwidth]{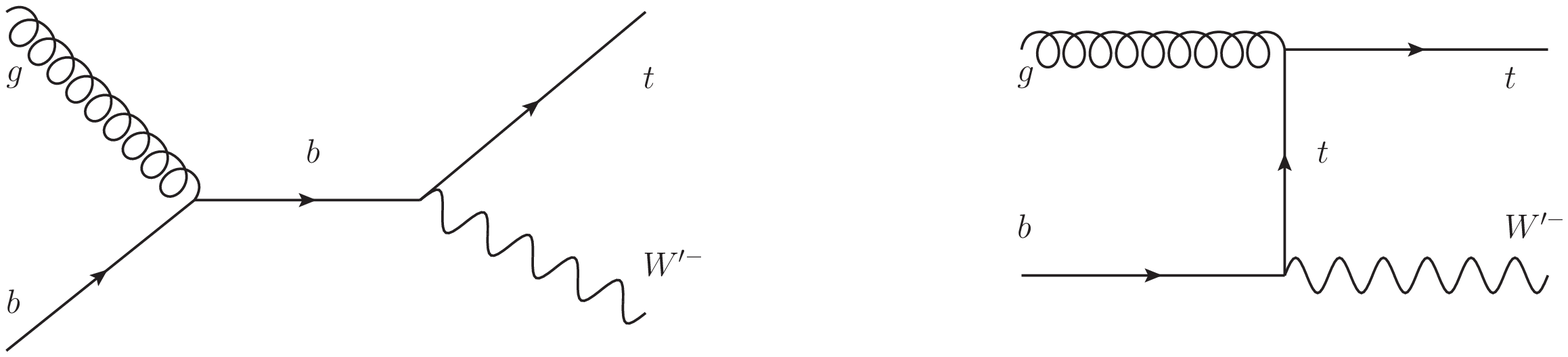}
\caption{Feynman diagrams for $gb\to tW^{\prime-}$ process at the tree level. }
\label{tw}
\end{figure}

\begin{figure}
\centering\includegraphics[width=0.40\textwidth]{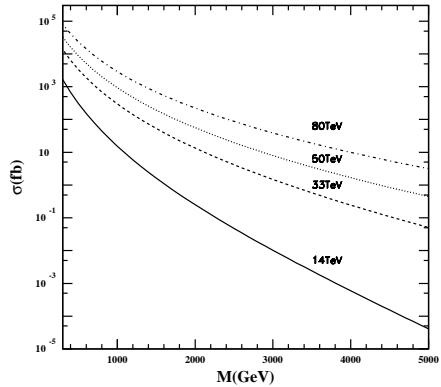}
\caption{The total cross section as a function of $W^\prime$ mass $M$ for $pp\to tW^{\prime -}$
process at the LHC for 14 TeV and 33 TeV, and future hadron collider for 50 TeV and 80 TeV.}
\label{fig:cs}
\end{figure}

The tree level predominant partonic sub-process for $tW^\prime$ production is the following process
\begin{equation}
g(p_1)+b(p_2)\to t(p_3)+W^{\prime -}(p_4),
\label{twchannel}
\end{equation}
as depicted in Fig.~\ref{tw}, where $p_i$ denotes the 4-momentum of the corresponding particle. The $g \bar b \to \bar t W^{\prime +}$ process is not referred in this paper for the similar character under the $CP-$invariance. The LHC leads to a large probability to study the $W^\prime$ for the larger gluon parton distribution than Tevatron.
The total cross section for the process $gb\to tW^{\prime{-}}$ can be expressed as
\begin{equation}
\sigma=\int f_{g}(x_{1})f_{b}(x_{2})\hat{\sigma}_{gb\to tW^{\prime^{-}}}(\hat s)dx_{1}dx_{2},
\label{cross section}
\end{equation}
where $f_{g}(x_{1})(f_{b}(x_{2}))$ is the parton distribution function (PDF) of gluon (quark),
$\hat s$ is the partonic center of mass energy squared, and $\hat{\sigma}$ is the partonic level cross section for $gb\to tW^{\prime{-}}$
process. To obtain the numerical results we set $V_{tb}=1$, $M_{W}=80.399$
GeV, and $m_{t}=173.1$ GeV. For PDF, we use CTEQ6L1\cite{Pumplin:2002vw}.
The total cross section for $pp\to tW^{\prime{-}}$ process is displayed in Fig.~\ref{fig:cs},
as a function of $M$ representing the mass of $W^{\prime -}_L$ and $W^{\prime -}_R$ at the LHC with 14 TeV and 33 TeV. In addition, we also give the results at the future hadron collider with 50 TeV and 80 TeV as a reference. There is no discrepancy in the cross section of left-handed $W^\prime$ production from right-handed. The $tW^{\prime -}$ cross section can be up to 15 (300) $fb$ for $M=1$ TeV with $\sqrt{s}=14$ (33) TeV at the LHC. While the observation of the $W^\prime$ signal at 14 TeV LHC for the mass heavier than 3.3 TeV is difficult in the $tW^{\prime -}$ associated production on the condition of the couplings chosen in this paper. The $W^\prime$ and top quark can not be observed directly at the LHC, thus we analyze the multi-jets plus lepton and missing transverse energy signal with different decay modes as the follows.

\subsection{$W^{\prime}\to tb$ channel for $tW^\prime$ production }
\label{tbchannel}

First we explore the $W^{\prime}\to tb$ channel for the $pp\to tW^{\prime -}$ process at the LHC with $\sqrt s=14$ and 33 TeV
as the following process
\begin{equation}
pp\to tW^{\prime-}\,\to\, t\bar{t}b\,\to\, bl^{+}\,+\, b\bar{b}jj\,+\,\met,\\
\label{eq:5jlplus}
\end{equation}
\begin{equation}
pp\to tW^{\prime-}\,\to\, t\bar{t}b\,\to\, bjj\,+\, b\bar{b}l^{-}\,+\,\met,
\label{eq:5jlminus}
\end{equation}
where the charged lepton is an excellent trigger for the event search.
Corresponding to process (\ref{eq:5jlplus}), the associated top quark semi-leptonically decays as $t\to bl^{+}\nu_{l}$ and the anti-top quark hadronically decays as $\bar{t}\to\bar{b}jj$. While in process (\ref{eq:5jlminus}), the decay modes of top and anti-top quark are exchanged.

To be more realistic, the simulation at the detector is performed
by smearing the leptons and jets energies according to the assumption
of the Gaussian resolution parameterization \begin{equation}
\frac{\delta(E)}{E}=\frac{a}{\sqrt{E}}\oplus b,\end{equation}
 where $\delta(E)/E$ is the energy resolution, $a$ is a sampling
term, $b$ is a constant term, and $\oplus$ denotes a sum in quadrature.
We take $a=5\%$, $b=0.55\%$ for leptons and $a=100\%$, $b=5\%$
for jets respectively\cite{atlas0901}.

\begin{figure}
\centering \includegraphics[width=0.30\textwidth]{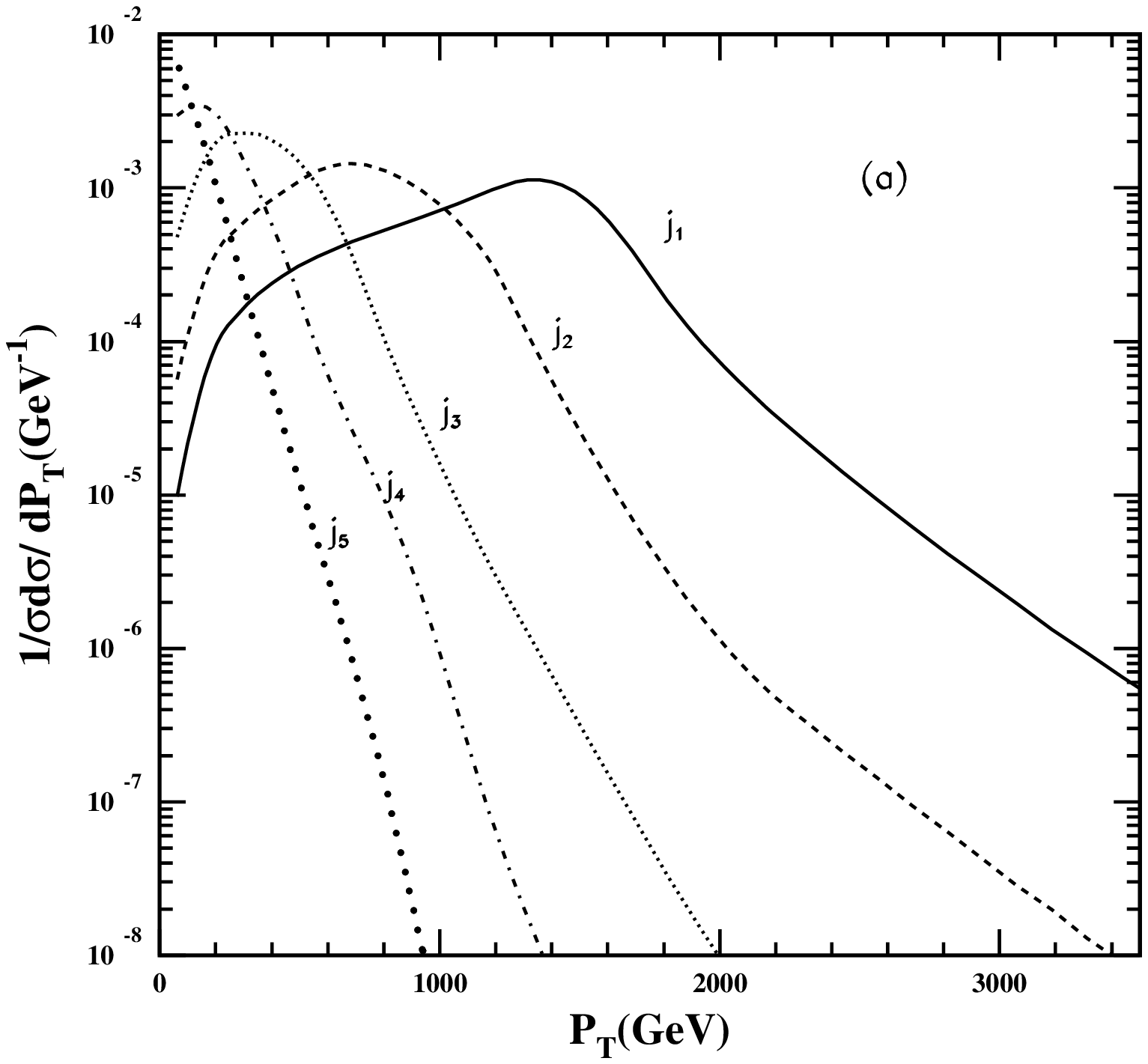}
\includegraphics[width=0.30\textwidth]{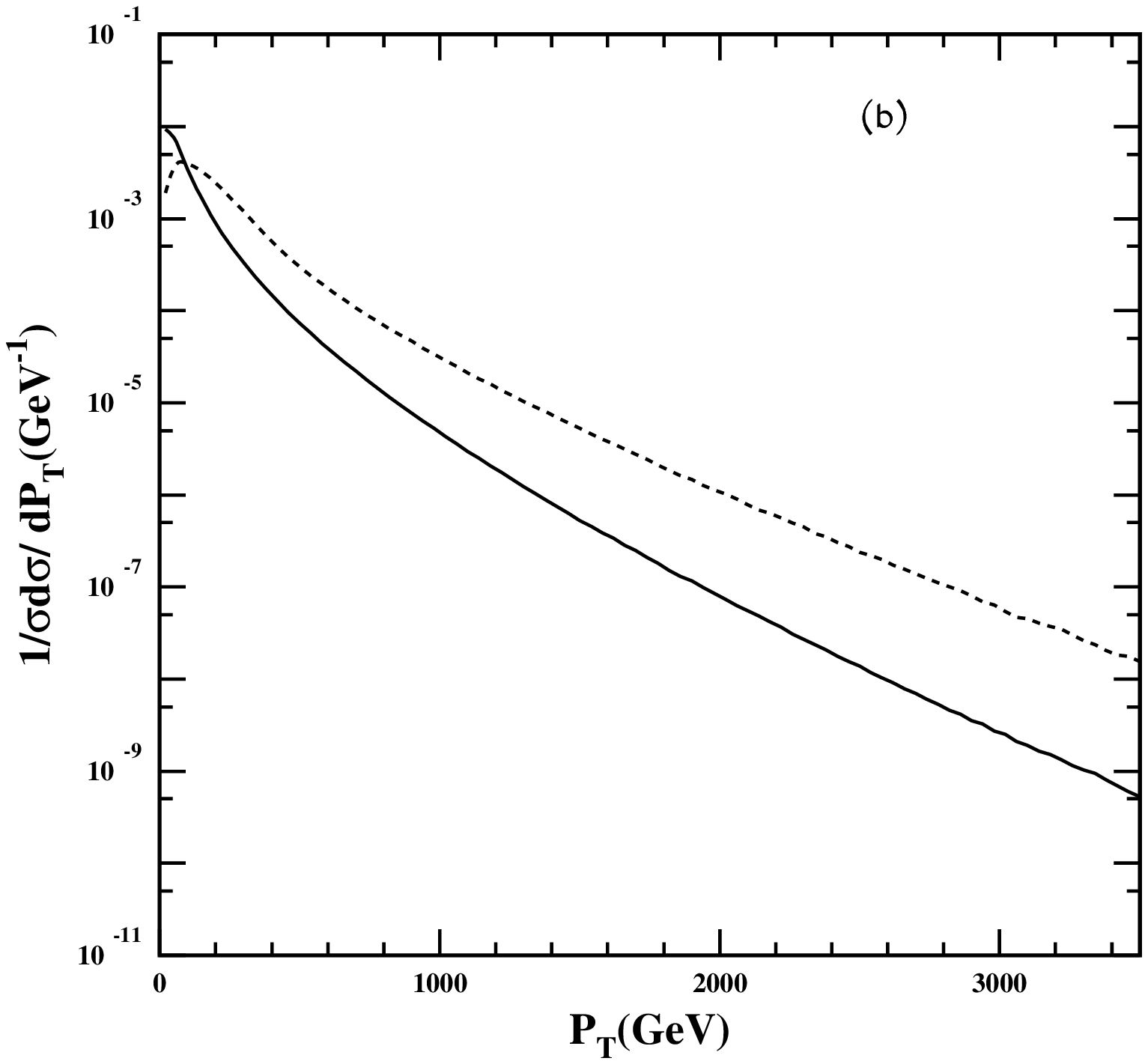}
\includegraphics[width=0.30\textwidth]{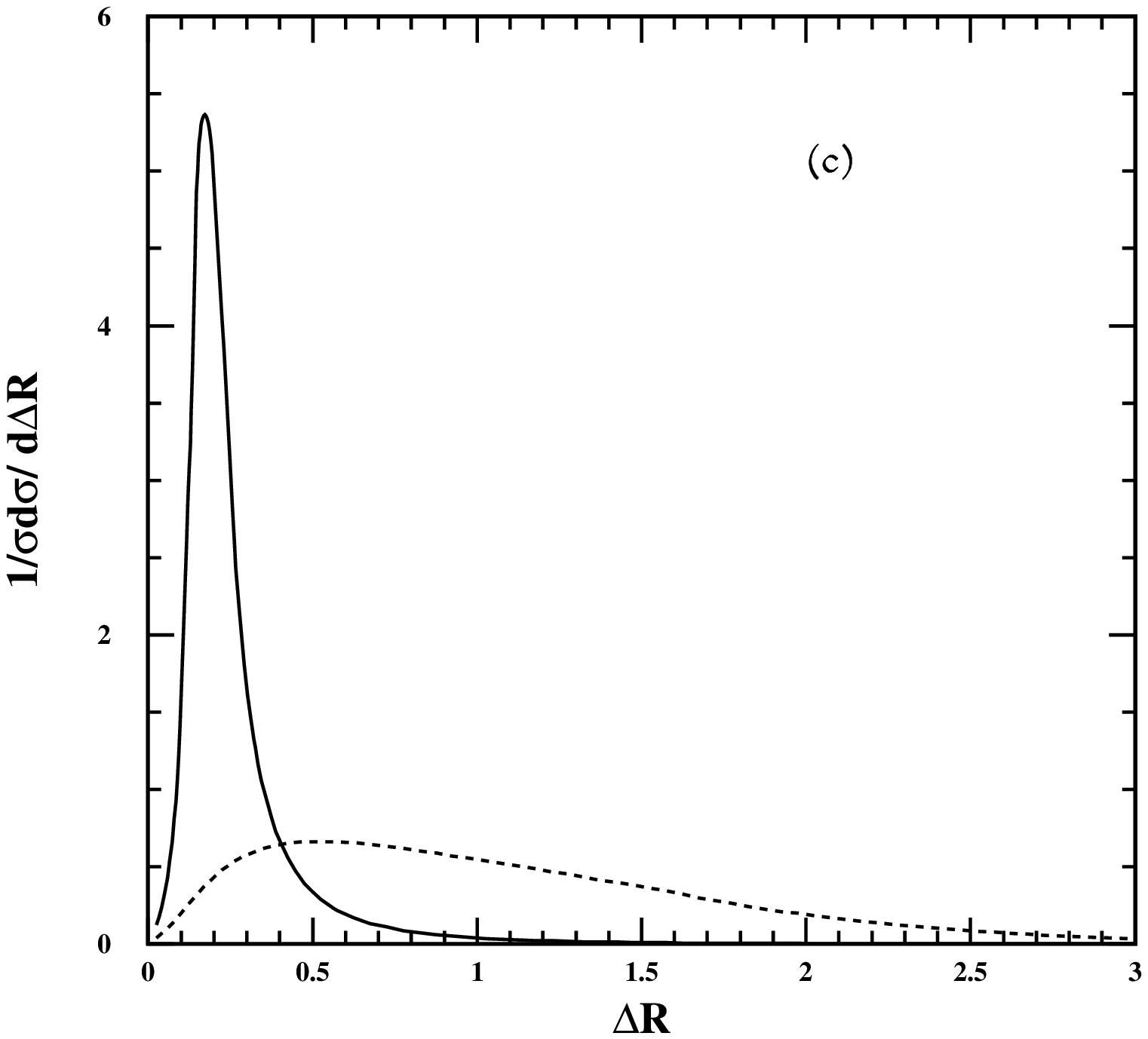}
\caption{(a) The normalized differential distributions with the transverse momentum of the jets ($j_1,j_2,j_3,j_4,j_5$)
with $p_{Tj_1}>p_{Tj_2}>p_{Tj_3}>p_{Tj_4}>p_{Tj_5}$ for $M=3$ TeV
in the process of $pp\to tW_L^{\prime-}\to bl^{+}+b\bar{b}jj+\met$ at $\sqrt{s}=33$ TeV.
(b) The same as (a) but for the charged lepton (solid line) and transverse missing energy $\met$ (dashed line).
(c) The minimal angular separation distributions between jets (solid line) and that between jets and the charged lepton (dashed line).}
\label{fig:pt}
\end{figure}

The transverse momentum distributions of jets and charged lepton are shown in Fig.~\ref{fig:pt} (a) and (b) as well as the missing transverse energy for the precess $pp\to tW^{\prime-}\to t\bar{t}b\to bl^{+}+b\bar{b}jj+\met$.
In order to identify the isolated jet (lepton), the angular distribution
between particle i and particle j is defined by
\begin{equation}
\Delta R_{ij}=\sqrt{\Delta\phi_{ij}^{2}+\Delta\eta_{ij}^{2}},
\end{equation}
where  $\Delta\phi_{i j}$ denotes the difference between the particles' azimuthal angle,
and  $\Delta\eta_{ij}$ the difference between the particles' rapidity.
In Fig.~\ref{fig:pt} (c), we display the differential distributions $\sigma^{-1}d\sigma/d\Delta R$ for $\Delta R=min(\Delta R_{ij})$.

The analysis of the whole process including the reconstruction of the intermediate resonances is propitious to select the $tW'$ production process from the substantial backgrounds. Theoretically, the top (anti-top) quark with semi-leptonical decay can be reconstructed by one of the five jets,
the charged lepton and the neutrino, while three of remaining jets can be used to reconstruct the anti-top (top) quark with hadronical decay. Although the  momentum of neutrino can not be directly recorded at the detector, one can resolve it by the kinematical constraint. The neutrinos' transverse momentum is determined by the sum of the observable particles' transverse momentum according to  the momentum conservation, and the longitudinal part can be solved through the on shell condition for the W-boson,
\begin{eqnarray}
&&{\bf p}_{\nu T}=-({\bf p}_{lT}+\sum_{j=1}^{5}{\bf p}_{jT}),\nonumber\\
&&m_{W}^{2}=(p_{\nu}+p_{l})^{2}.
\label{neutrino}
\end{eqnarray}
Once the neutrino's momentum reconstructed, we can reconstruct the top or anti-top quark invariant mass
\begin{equation}
M_{jl\nu}^{2}=(p_{l}+p_{\nu}+p_{j})^{2},
\end{equation}
where j refers to any one of the five jets that makes the $M_{jl\nu}$ be closest to the top quark mass.
Based on the above discussion, we employ the basic cuts to outstand the $tW'$ process as

\begin{itemize}
\item Cut $A_{1}$:
\begin{flalign}
&{\rm For~~ 14~~and~~33~~ TeV}\left\{\begin{array}{ll}
& p_{lT}>50~{\rm GeV},~~~~p_{jT}>50~{\rm GeV},~~~~\met>50~{\rm GeV},\nonumber \\
& |\eta_{l}|<2.5,~~~~|\eta_{j}|<2.5,~~~~\Delta R_{jj(lj)}>0.4,\nonumber \\
& |M_{j_{a}l\nu}-m_{t}|\leq30~{\rm GeV},~~~~|M_{j_{b}j_{c}j_{d}}-m_{t}|\leq30~{\rm GeV},\nonumber\\
&|M_{j_{b}j_{c}}-m_{W}|\leq10~{\rm GeV}.
\end{array} \right.&
\end{flalign}
\end{itemize}

\begin{figure}
\centering \includegraphics[width=0.40\textwidth]{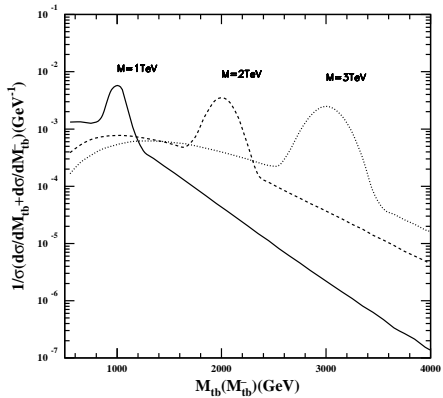}
\caption{The distributions of $1/\sigma(d\sigma/dM_{tb}+d\sigma/dM_{\bar{t}b})$ with respect to
the invariant mass reconstructed by top (anti-top) quark and the remaining jet
for $M=1,~2,~3$ TeV  for the process of
$pp\to tW_L^{\prime-}\to bl^{+}+b\bar{b}jj+\met$ after cut $A_{1}$ at the LHC for 33 TeV.}
\label{fig:mtb}
\end{figure}

Then the $W^\prime$ mass peak can be reconstructed through the momentum of all the particles except the ones used to reconstruct the associated top or anti-top quark. The distributions of $1/\sigma(d\sigma/dM_{tb}+d\sigma/dM_{\bar{t}b})$ are shown
in Fig.~\ref{fig:mtb}. The resonance peak is significant in the invariant mass distribution of $tb$, which will be a direct signal in the search of $W^\prime$. In order to further purify the signal, we require the following cut
\begin{itemize}
\item Cut $A_{2}$:$~~|M_{jj_{b}j_{c}j_{d}}-M|\leq10\%\, M$
or $|M_{jj_{a}l\nu}-M|\leq10\%\, M$,
\end{itemize}
together with the remaining jet that not be used to reconstruct top or anti-top quark tagged as a b-jet. The $b$-tagging efficiency is assumed to be 60\% while the miss-tagging efficiency of a light jet as
a $b$ jet is taken as transverse momentum dependent \cite{atlas0901}:
\begin{equation}
\epsilon_l =\left\{ \begin{array}{ll}
\displaystyle{\frac{1}{150}}, &P_T< 100\, {\rm GeV},\\
\displaystyle{\frac{1}{450}}\,\left(\frac{P_T}{25\,{\rm GeV}}-1\right), &100 \,{\rm GeV}\leq P_T<250\, {\rm GeV},\\
\displaystyle{\frac{1}{50}}, & P_T\geq 250\, {\rm GeV}.
\end{array} \right.
\end{equation}

\begin{figure}
\centering
\includegraphics[width=0.40\textwidth]{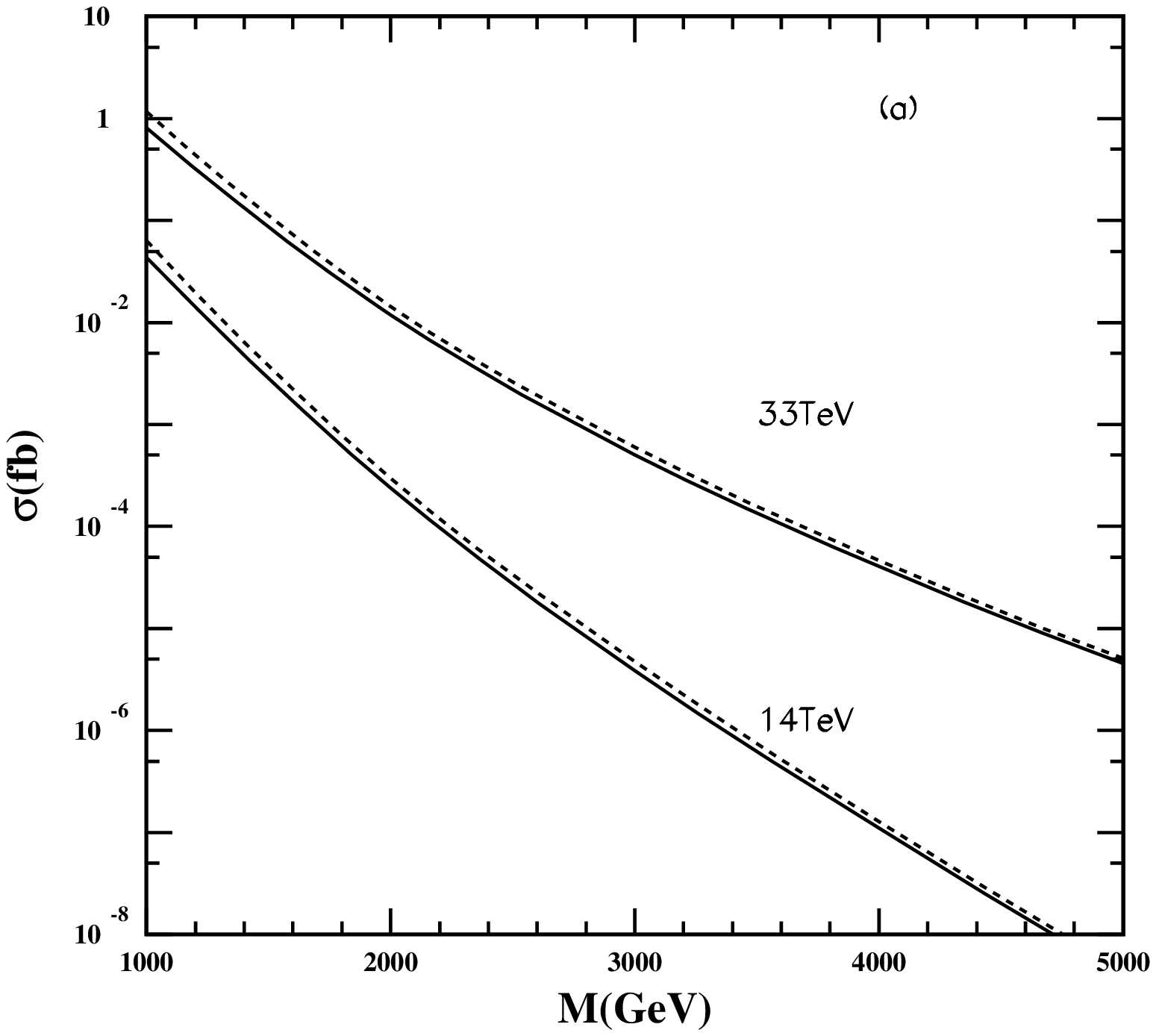}
\includegraphics[width=0.40\textwidth]{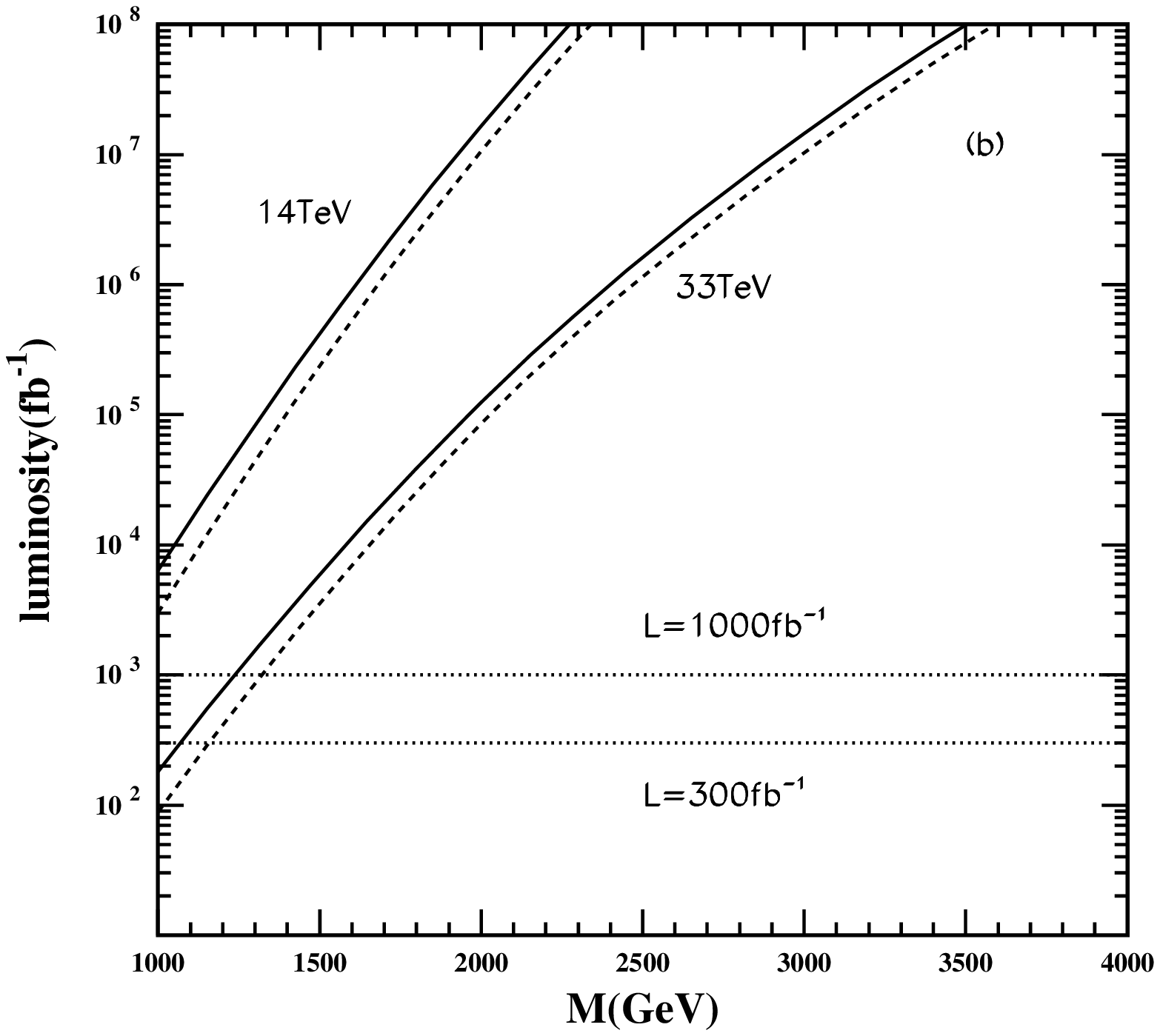}
\caption{(a) The total cross section  as a function of
the charged gauge bosons mass $M$ at $\sqrt{s}=14$ and $33$ TeV
for the process of $pp\to tW^{\prime-}\to 5{\rm jets}+l^{+}+\met$  after all the cuts.
(b) The integral luminosity needed at the LHC for $\sqrt{s}=14$ and $33$ TeV
at $S/\sqrt{B}=3\sigma$ sensitivity. The solid and dashed lines stand for $W_L^{\prime-}$ and $W_R^{\prime-}$ respectively, and the two straight dotted lines present the luminosity of $300$ and $1000~fb^{-1}$.}
\label{fig:lum-tb1}
\end{figure}

\begin{figure}
\centering
\includegraphics[width=0.40\textwidth]{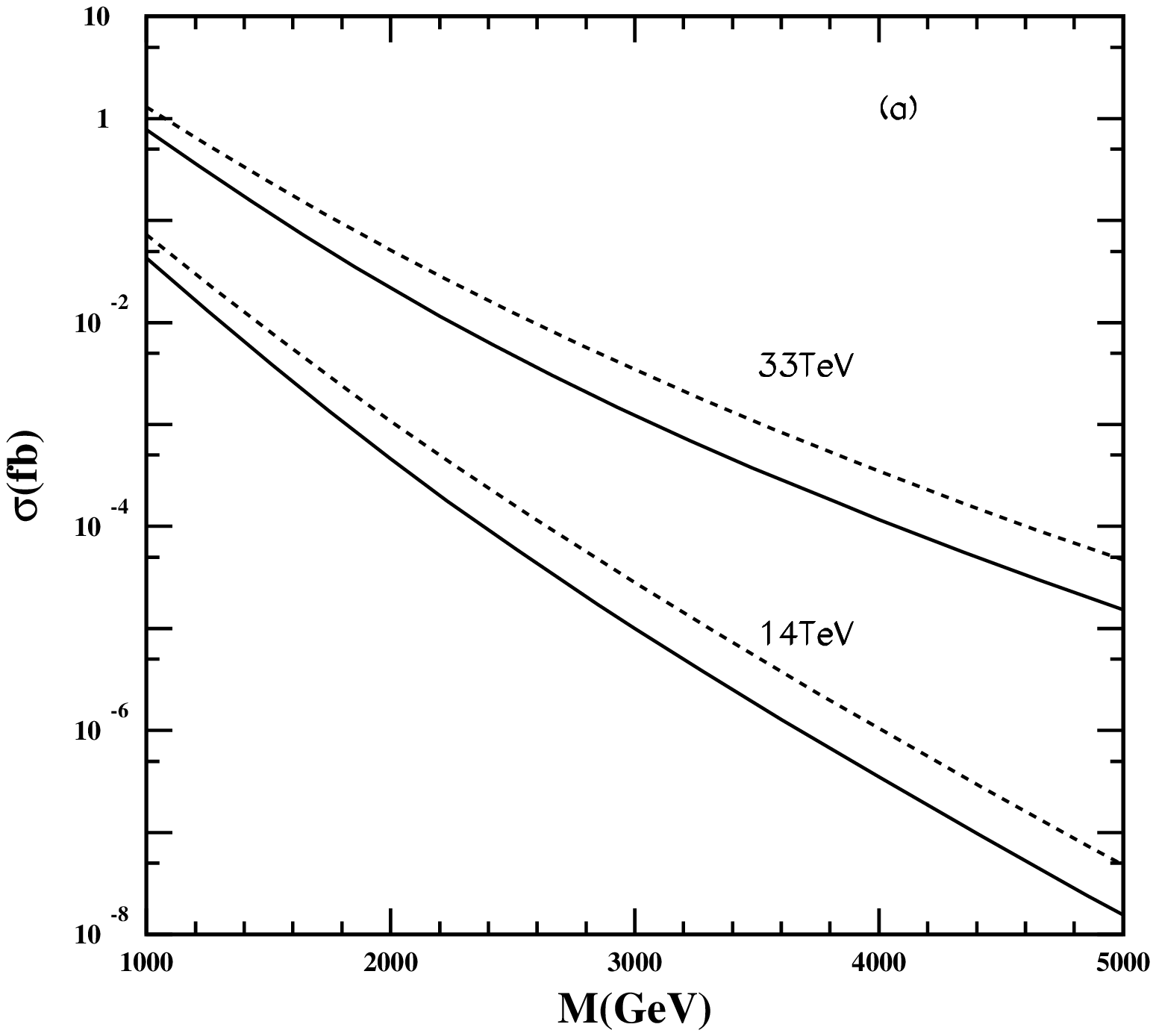}
\includegraphics[width=0.40\textwidth]{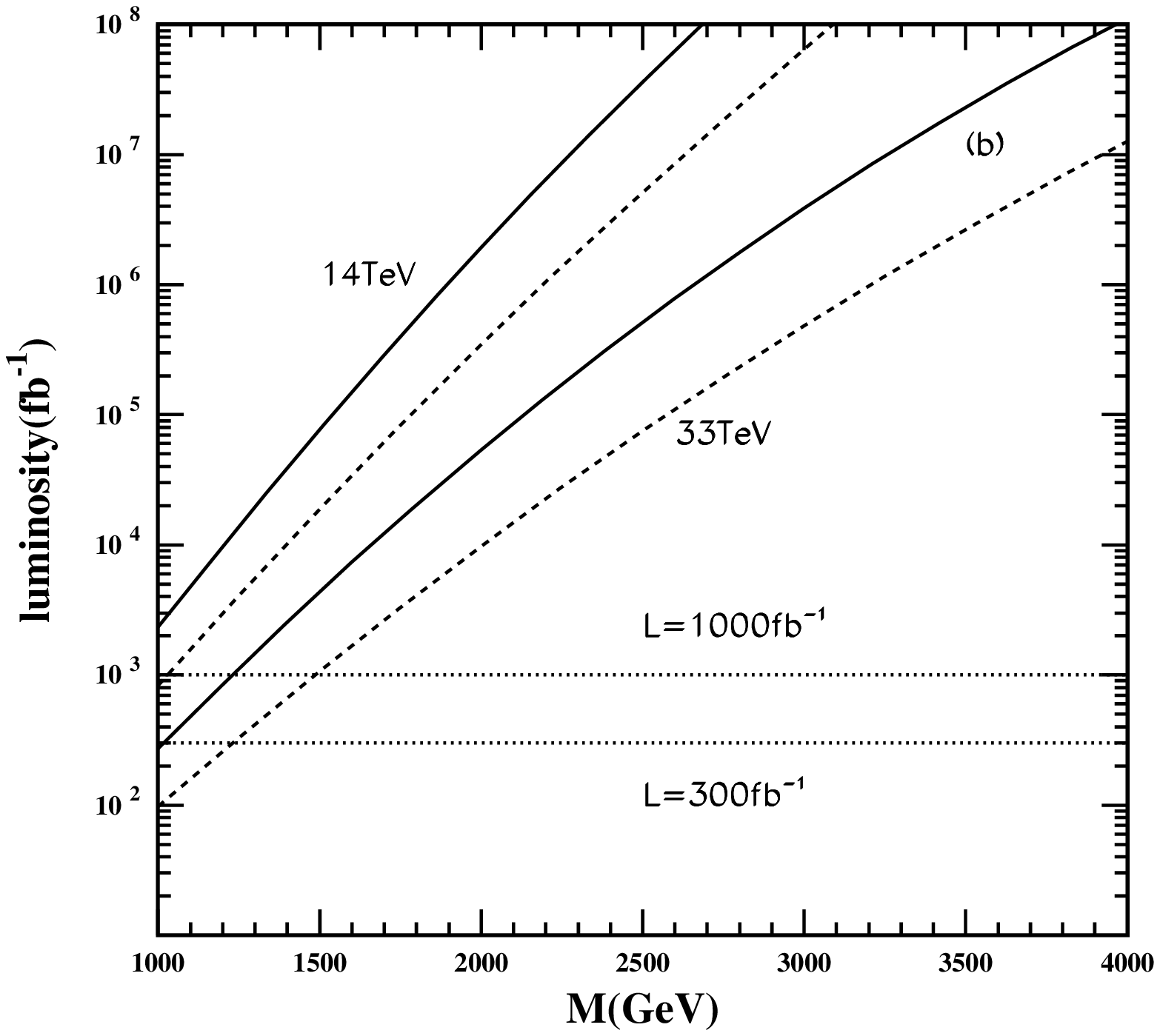}
\caption{(a) The total cross section  as a function of
the charged gauge bosons mass $M$ at $\sqrt{s}=14$ and $33$ TeV
for the process of $pp\to tW^{\prime-}\to 5{\rm jets}+l^{-}+\met$  after all the cuts.
(b) The integral luminosity needed at the LHC for $\sqrt{s}=14$ and $33$ TeV
at $S/\sqrt{B}=3\sigma$ sensitivity. The solid and dashed lines stand for $W_L^{\prime-}$ and $W_R^{\prime-}$ respectively, and the two straight dotted lines present the luminosity of $300$ and $1000~fb^{-1}$.}
\label{fig:lum-tb2}
\end{figure}

The cross section of process (\ref{eq:5jlplus}) at the LHC with $\sqrt{s}=14 $ and $33$ TeV after Cut $A_{1}$ and $A_{2}$ are displayed as a function of $M$ in Fig.~\ref{fig:lum-tb1} (a), which deriving from $W^\prime _R$  is larger than $W^\prime _L$ due to the discrepancy of the branching fraction $W^\prime \to tb$. If the mass of $W^\prime$ is larger than 1 TeV, it will be negative to observe $W^\prime$ from the process at the 14 TeV LHC, while the cross section can be enhanced more than one order of magnitude at 33 TeV. Considering the effect from  the background of $t\bar t j$, we give the integral luminosity needed at the LHC with $S/\sqrt{B}=3\sigma$ for various $M$ in Fig~.\ref{fig:lum-tb1} (b). The result shows that the $W^\prime$ with mass below 1 TeV can be observed from the process (\ref{eq:5jlplus}) at the LHC with  $\sqrt{s}=33 $ TeV. For a $W^\prime$ with mass up to 3 TeV, only with the sizable couplings it can be detected in the $tW'$ associated production via the $W^\prime \to tb$ decay
at the future LHC. It makes no difference to the results for the process (\ref{eq:5jlminus}) as displayed in Fig.~\ref{fig:lum-tb2}.

\subsection{$W^\prime \to q\bar q'$ channel for $tW^\prime$ production}

\begin{figure}
\centering \includegraphics[width=0.30\textwidth]{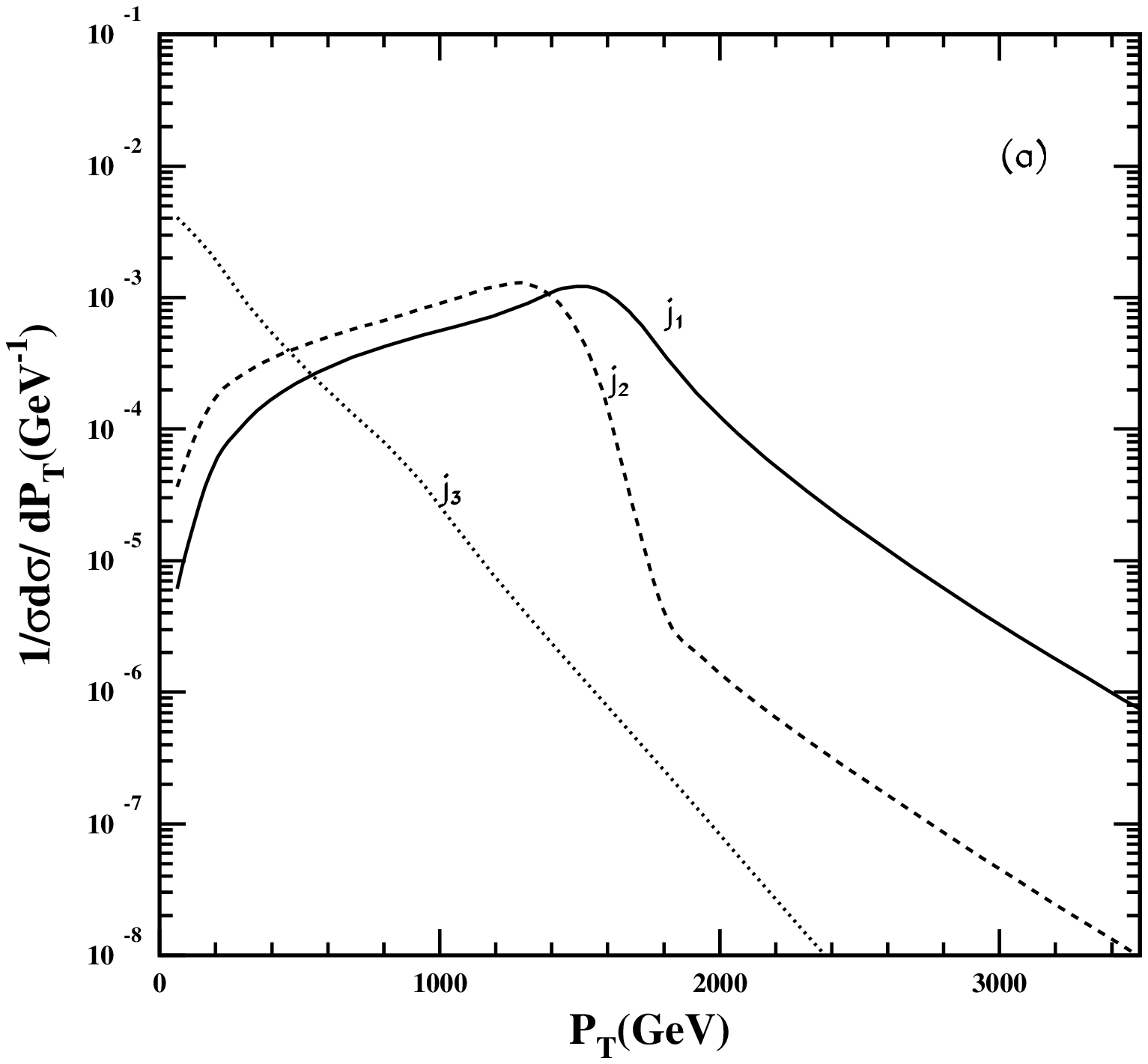}
\includegraphics[width=0.30\textwidth]{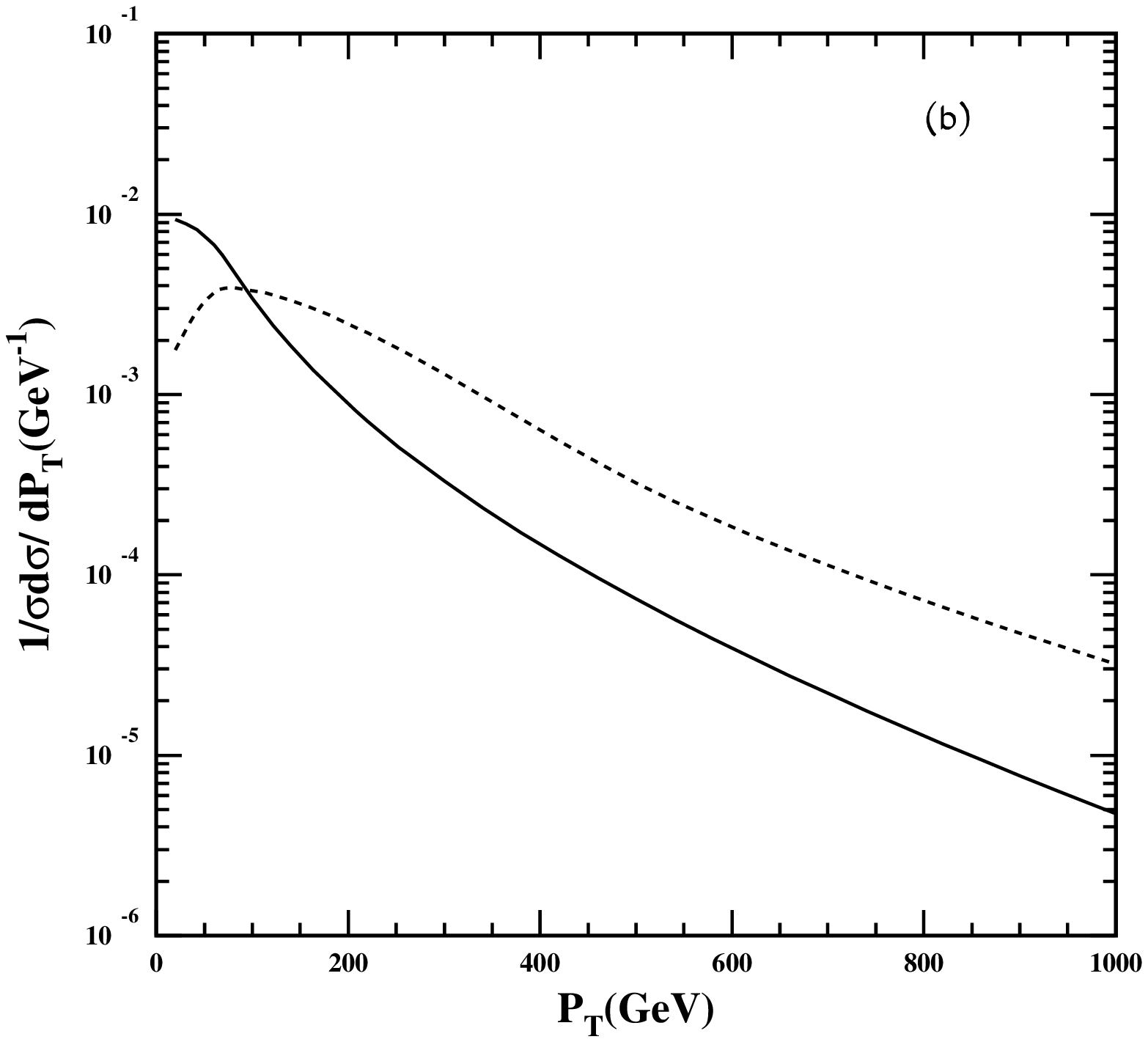}
\includegraphics[width=0.30\textwidth]{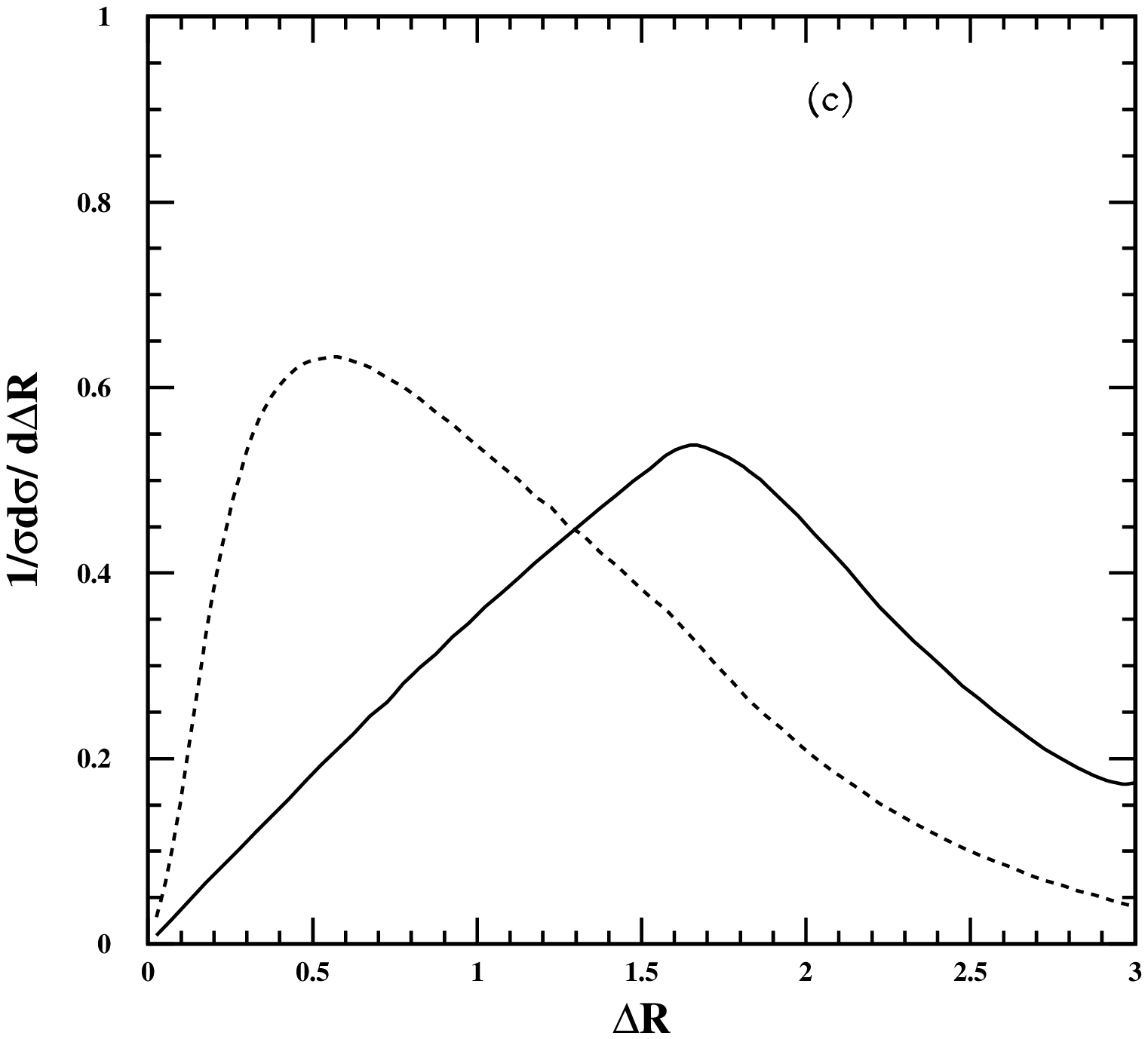}
\caption{(a) The normalized differential distributions with the transverse momentum of the jets ($j_1,j_2,j_3$)
with $p_{Tj_1}>p_{Tj_2}>p_{Tj_3}$ for $M=3$ TeV in the process of $pp\to tW_L^{\prime-}\to l^{+}+3jets+\met$ at $\sqrt{s}=33$ TeV.
(b) The same as (a) but for the charged lepton (solid line) and transverse missing energy $\met$ (dashed line).
(c) The minimal angular separation distributions between jets (solid line) and that between jets and the charged lepton (dashed line).}
\label{fig:pt-qq}
\end{figure}

According to Fig. \ref{fig:ratio}, the search of $W^\prime$ is in favor of $W^\prime \to q\bar q'$ decay modes for the large branch fraction. Then we focus on the following process
\begin{equation}
pp\to tW^{\prime -} \to bW^+ W^{\prime -}\to bl^{+}+jj+\met,
\label{qqdeay}
\end{equation}
where the charged lepton is from the associated top quark. The  differential distributions with the transverse momentum of the three jets and charged lepton are shown in Fig.~\ref{fig:pt-qq} (a) and (b) as well as the missing transverse energy. The first two jets with the largest transverse momentum mostly derive from $W^\prime$ so that two peaks appear nearby 1.5 TeV which is half of the $W^\prime$ mass. The normalized differential distributions with $\Delta R$  which are displayed in  Fig.~\ref{fig:pt-qq} (c) for $pp\to tW^{\prime-}\to bl^{+}+jj+\met$ process is broader than that in processes (\ref{eq:5jlplus}) and (\ref{eq:5jlminus}).

We use the same methods as in Sec. III A to obtain the momentum of neutrino and reconstruct the intermediate resonances.
The following relation is adopted,
\begin{eqnarray}
&&{\bf p}_{\nu T}=-({\bf p}_{lT}+\sum_{j=1}^{3}{\bf p}_{jT}),\nonumber\\\
&&m_{W}^{2}=(p_{\nu}+p_{l})^{2},\nonumber\\\
&&M_{jl\nu}^{2}=(p_{l}^{2}+p_{\nu}+p_{j})^{2},
\end{eqnarray}
with j named $j_a$ refers to any one of the three jets which makes $M_{jl\nu}$ be closest to the top quark mass and $j_{b(c)}$ for the left two.

Considering the unlike transverse momentum distributions of jet with the center of mass energy at 14 and 33 TeV, we employ the basic cuts as
\begin{itemize}
\item Cut $B_{1}$:
\begin{flalign}
&{\rm For~~ 14~~ TeV}\left\{\begin{array}{ll}
&p_{j_{1}T}>200~{\rm GeV},~~~~p_{j_{2}T}>100~{\rm GeV},~~~~p_{j_{3}T}>20~~{\rm GeV},\nonumber \\
&p_{lT}>20~~{\rm GeV},~~~~~\met>20~{\rm GeV},\nonumber \\
&|\eta_{l}|<2.5,~~~~|\eta_{j}|<2.5,~~~~~\Delta R_{jj(lj)}>0.4,\nonumber \\
&|M_{j_{a}l\nu}-m_{t}|\leq30~{\rm GeV}.\end{array} \right.&\\
&{\rm For~~ 33~~ TeV}\left\{\begin{array}{ll}
&p_{j_{1}T}>550~{\rm GeV},~~~~p_{j_{2}T}>550~{\rm GeV},~~~~p_{j_{3}T}>20~~{\rm GeV},\nonumber \\
&p_{lT}>20~~{\rm GeV},~~~~~\met>20~{\rm GeV},\nonumber \\
&|\eta_{l}|<2.5,~~~~|\eta_{j}|<2.5,~~~~~\Delta R_{jj(lj)}>0.4,\nonumber \\
&|M_{j_{a}l\nu}-m_{t}|\leq30~{\rm GeV}.\end{array} \right.&
\end{flalign}
\end{itemize}

\begin{figure}
\centering \includegraphics[width=0.40\textwidth]{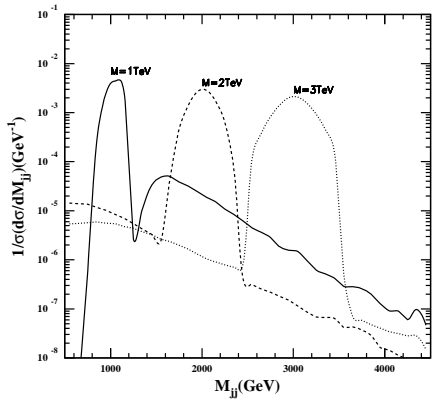}
\caption{The distributions of $1/\sigma d\sigma/dM_{jj}$ with respect to
the invariant mass reconstructed by the remaining  two jets
for $M=1,~2,~3$ TeV for the process of
$pp\to tW_L^{\prime-}\to 3{\rm jets}+l^{+}+\met$  after cut $B_{1}$ at the LHC for 33 TeV.}
\label{fig:mqq}
\end{figure}

Once the jet derived from top quark is confirmed by the reconstruction of top quark, the two remaining jets  are absolutely from the heavy charged gauge boson $W^\prime$. As presented in Fig.~\ref{fig:mqq}, the resonance peak is obvious around the $W^\prime$ mass in the differential distribution with $M_{j_bj_c}$.
Hence the further cut is
\begin{itemize}
\item Cut $B_{2}$:$~~~|M_{j_bj_c}-M|\leq10\%\, M $.
\end{itemize}
Comparing the signal to the main SM backgrounds $Wjjj$, $WWj$ and $WZj$, one can find that b-tagging can help to purify the signal, so we require the jet used to reconstruct top quark to be a b-jet.

\begin{figure}
\centering
\includegraphics[width=0.4\textwidth]{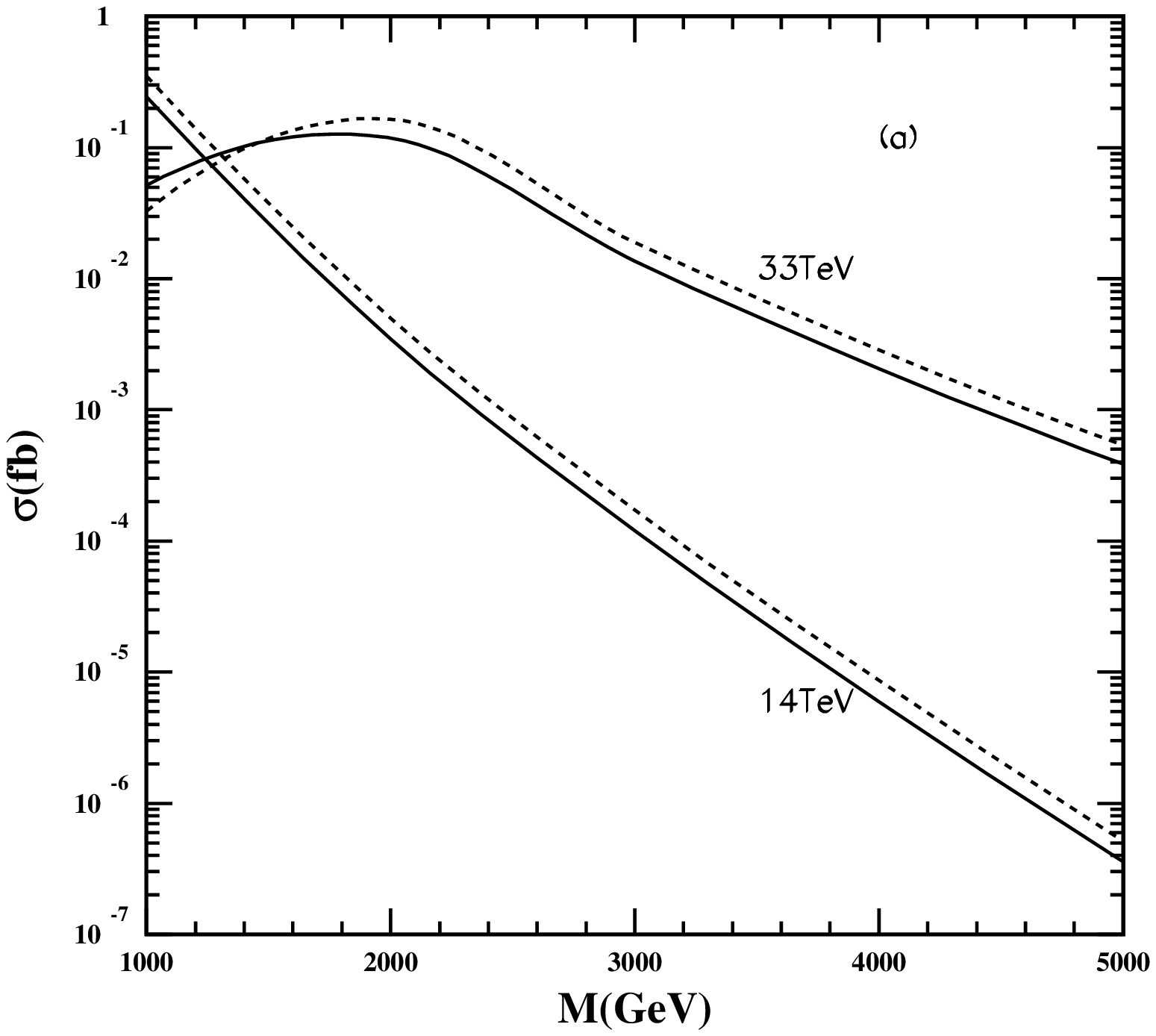}
\includegraphics[width=0.4\textwidth]{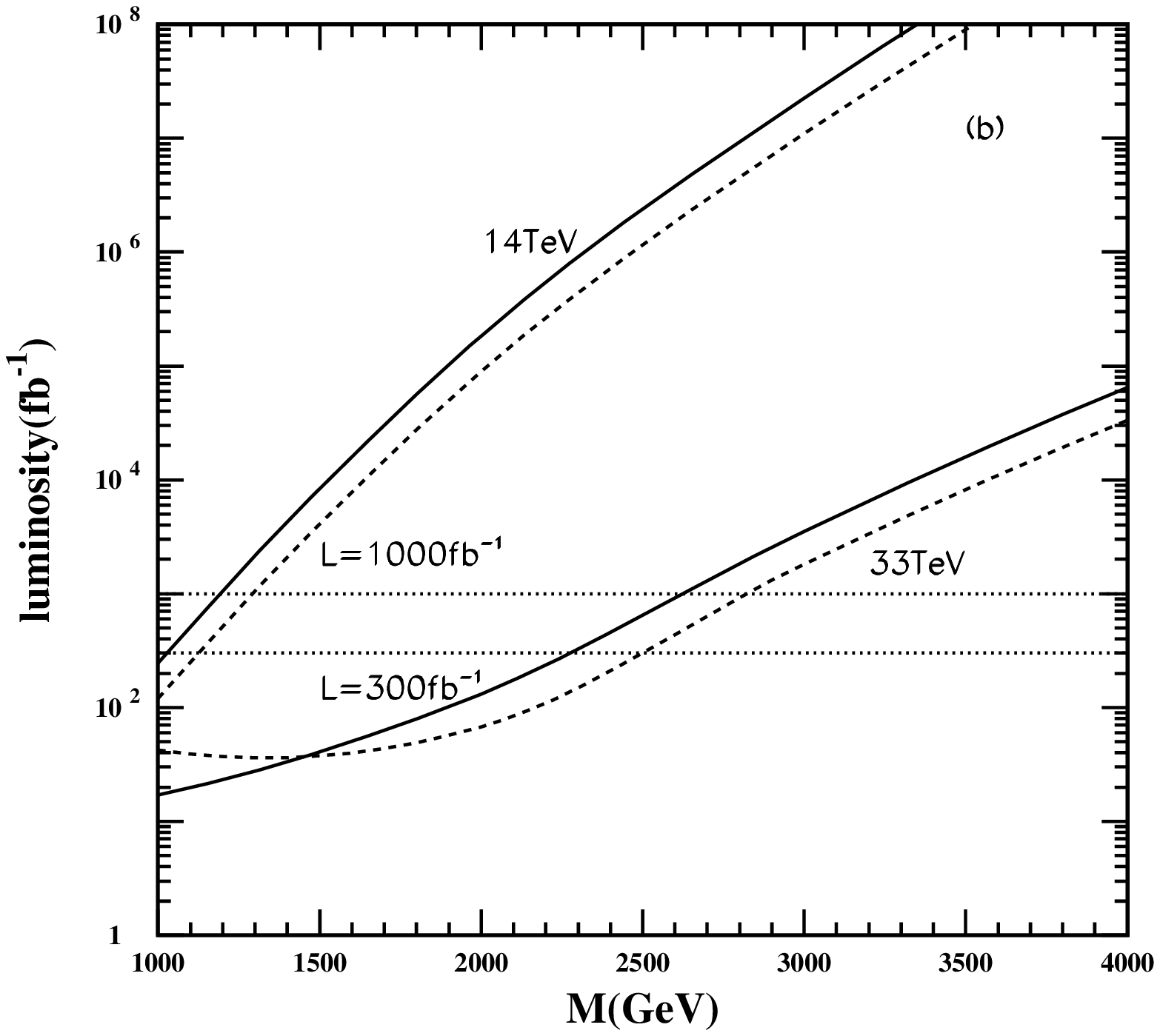}
\caption{(a) The total cross section  as a function of
the charged gauge bosons mass $M$ at $\sqrt{s}=14$ and $33$ TeV
for the process of $pp\to tW^{\prime-}\to 3{\rm jets}+l^{+}+\met$.
(b) The integral luminosity needed at the LHC for $\sqrt{s}=14$ and $33$ TeV
at $S/\sqrt{B}=3\sigma$ sensitivity. The solid and dashed lines stand for $W_L^{\prime-}$ and $W_R^{\prime-}$ respectively, and the two straight dotted lines present the luminosity of $300$ and $1000~fb^{-1}$.  }
\label{fig:lum-qq}
\end{figure}

We show the total cross section after all the cuts for the signal process (\ref{qqdeay}) at the LHC
with  $14$ TeV and $33$ TeV in Fig.~\ref{fig:lum-qq} (a), as well as the integral luminosity as a function of $M$  assuming the significance is $3\sigma$ in Fig.~\ref{fig:lum-qq} (b).
The cross section is up to 0.003 (0.005) $fb$ for a mass of 2 TeV $W^{\prime}_{L}$ ($W^{\prime}_{R}$)  with 14 TeV, as well as 0.118 (0.165) $fb$ with 33 TeV,  while it is obviously suppressed by the strictly kinematic cuts for the $W^\prime$ mass lighter than 2 TeV with 33 TeV.
The main backgrounds with the same detector signal including $tW$, $Wjjj$, $WWj$ and $WZj$ are simulated by MadEvent  program~\cite{MadEvent}.  One can find that the luminosity need to be up to $10^5 fb^{-1}$ if the charged bosons with $M=2$ TeV can be detected at $14$ TeV.
However, there are about forty events can be detected for a  $M=2.8$ TeV $W^\prime$ with the luminosity of $1000 fb^{-1}$ at $33$ TeV.

Once a heavy charged boson $W^\prime$ is discovered at the LHC, it will be imperative to determine
its chiral couplings to SM fermions.
Accounting for different chiral couplings between $W^\prime _Lq\bar q'$ and $W^\prime _Rq\bar q'$,
we investigate the charged lepton angular distribution which depends on the chiral couplings of $W^\prime$ to light quarks. The chirality of the $W^\prime$ coupling to the light quarks can be translated to the
angular distribution of the charged lepton and a forward-backward asymmetry is defined as follows
\begin{eqnarray}
&&\frac{1}{\sigma}\,\frac{d\sigma}{d\cos\theta^{*}}\,=\,\frac{1}{2}\,[1\,+\, A_{FB}\cos\theta^{*}],~~~~~~
A_{FB}=\frac{\sigma(cos\theta^{*}\geq0)-\sigma(cos\theta^{*}<0)}{\sigma(cos\theta^{*}\geq0)+\sigma(cos\theta^{*}<0)}
\label{eqfbeq}
\end{eqnarray}
with
\begin{equation}
\cos\theta^{*}=\frac{{\bf p}_{l^{+}}^{*}\cdot{\bf p}_{t}^{*}}{|{\bf p}_{l^{+}}^{*}||{\bf p}_{t}^{*}|}  .
\end{equation}
Here ${\bf p}_{l^{+}}^{*}$ is the 3-momentum of charged lepton in
the top quark rest frame, and ${\bf p}_{t}^{*}$ is the 3-momentum of the
top quark in $tW^{\prime -}$ center of mass frame.

\begin{figure}[htbp]
\centering
\includegraphics[width=0.4\textwidth]{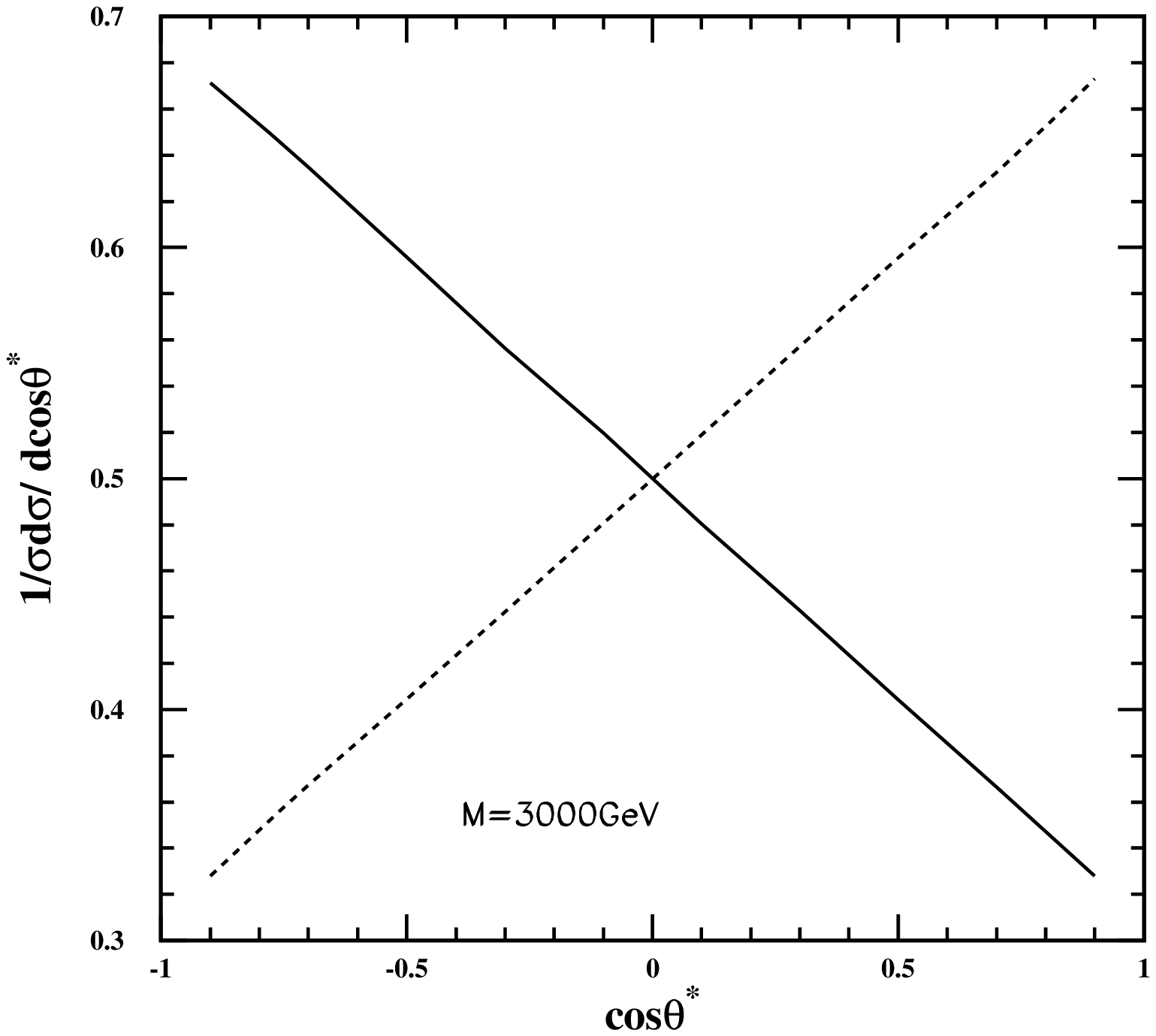}
\includegraphics[width=0.4\textwidth]{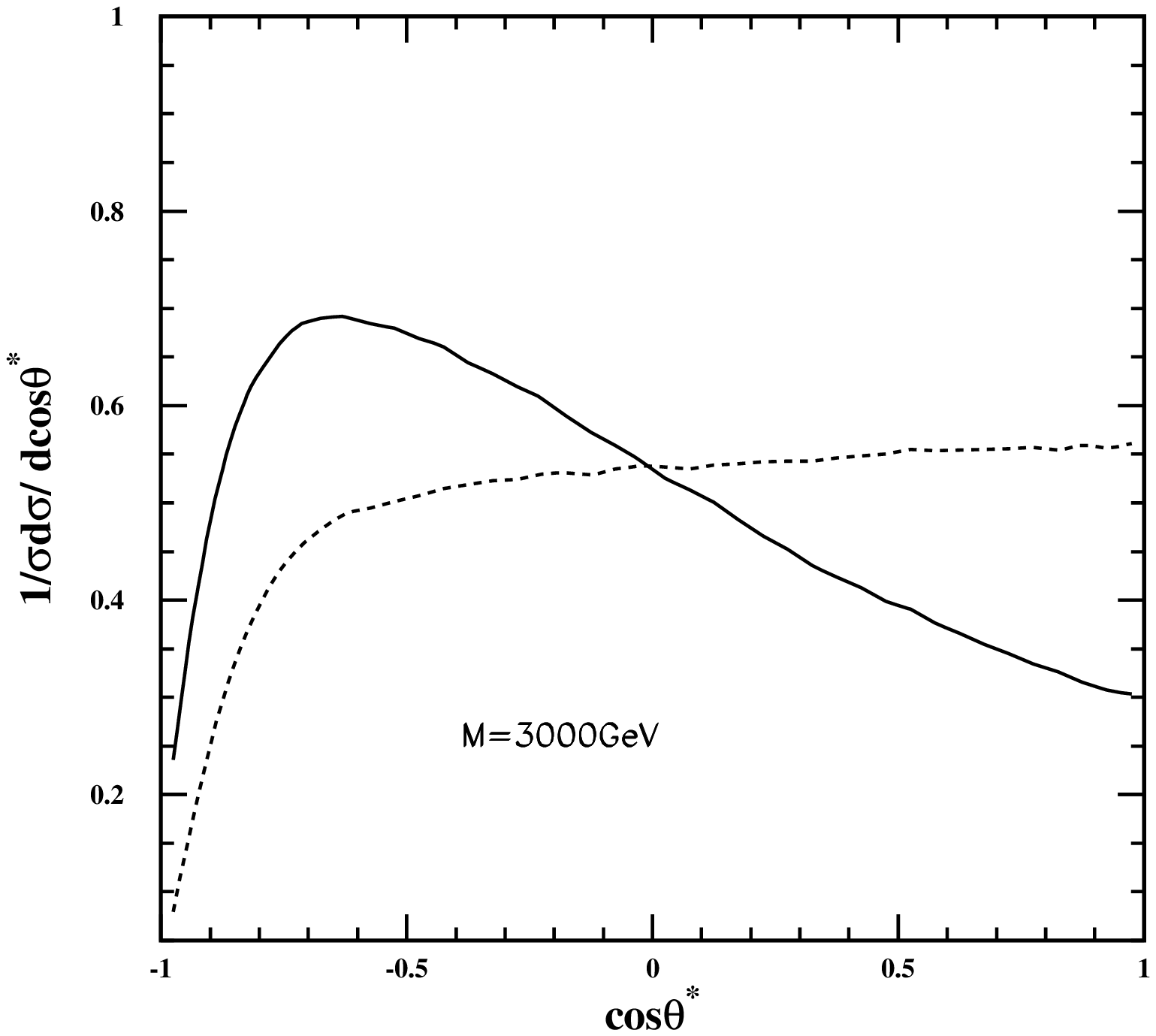}
\caption{The angular distributions of the
charged lepton for $M=3$ TeV with (left) and without (right) cuts
at $\sqrt{s}=33$ TeV for the process of $pp\to tW^{\prime-}\to 3{\rm jets}+l^{+}+\met$. The solid line and dashed line stand for $W^{\prime-}_L$ and $W^{\prime-}_R$, respectively.}
\label{fig:cos-3jl+}
\end{figure}

The charged lepton angular distributions with respect to $cos\theta^{*}$ before and after cuts are displayed in Fig.~\ref{fig:cos-3jl+} corresponding to the process (\ref{qqdeay}) at 33 TeV LHC.
The result shows that most charged leptons moving against the direction of the top quark for the left-handed type interaction, while it leads to the inverse tendency for the right-handed type interaction. Thus we can separate the hemisphere of top quark direction from the opposite hemisphere according to $\cos\theta^* \ge 0$ or $\cos\theta^* < 0$, then the forward-backward asymmetry leads to inverse sign which is listed in Table.~\ref{tab:asyqq}. Although the distributions after the cuts in the $cos\theta^*=-1$ region are severely distorted by the acceptance cuts,
due to the charged leptons moving against the top quark direction carry less energy than those in the remaining region, the forward-backward asymmetry with $-0.5<\cos\theta<0.5$ is also an excellent characteristic quantity to distinguish $W^{\prime}_L$ from $W^{\prime}_R$.

\begin{table}[htbp]
\begin{centering}\scalebox{0.9}{$
\begin{tabular}{|c||c|c|c||c|c|c|}
\hline
 $A_{FB}$&\multicolumn{3}{c||}{before cuts }& \multicolumn{3}{c|}{after cuts} \tabularnewline
\hline
$M$ (TeV) &1&2&3&1&2&3 \tabularnewline
\hline
$W^\prime_L$ &-0.147  & -0.174 &  -0.191&-0.426&-0.116 &-0.139\tabularnewline
\hline
$W^\prime_R$&0.147   &  0.174 &   0.191&0.035&0.020&0.016 \tabularnewline
\hline
\end{tabular}$}\caption{The forward-backward asymmetry $A_{FB}$
for $pp\to tW^{\prime-}\to 3{\rm jets}+l^{+}+\met$ at the LHC with $\sqrt{s}=33$ TeV before
and after the cuts.}
\label{tab:asyqq}
\par\end{centering}
\centering{}
\end{table}

\subsection{$W^\prime\to l\nu$ channel for $tW^\prime$ production}

The leptonic decay modes of $W^\prime$ depend on the lepton spectrum and the flavor mixing in the given model.
Although the right-handed $W^\prime$ boson can couple to a charged lepton and right-handed neutrino in some new physics models, the decay modes of $W^\prime _R\to l \nu_R$ is not considered due to the mass of $\nu_R$ larger than $W^\prime$ or the different signal comparing to $W^\prime _L\to l \nu$. In this section, we focus on the  process
\begin{equation}
pp\to tW^{\prime -}_L \to bW^+ W^{\prime -}\to bjj+l^{-}+\met
\label{lvchannel}
\end{equation}
with $W^{\prime -} \to l^- \nu$ decay mode, which provides the charged lepton with large transverse momentum as an excellent trigger at the LHC.

\begin{figure}
\centering \includegraphics[width=0.30\textwidth]{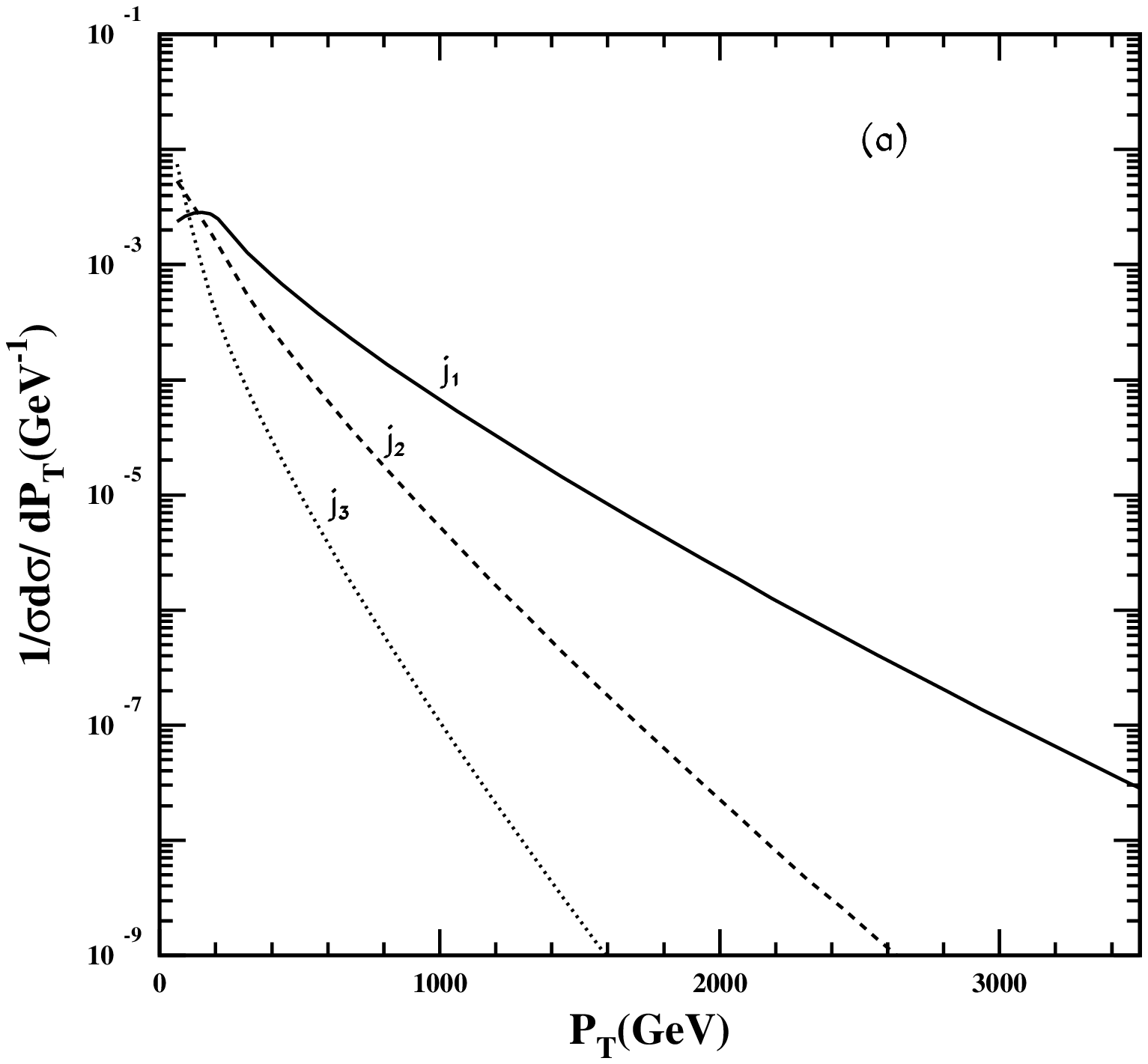}
\includegraphics[width=0.30\textwidth]{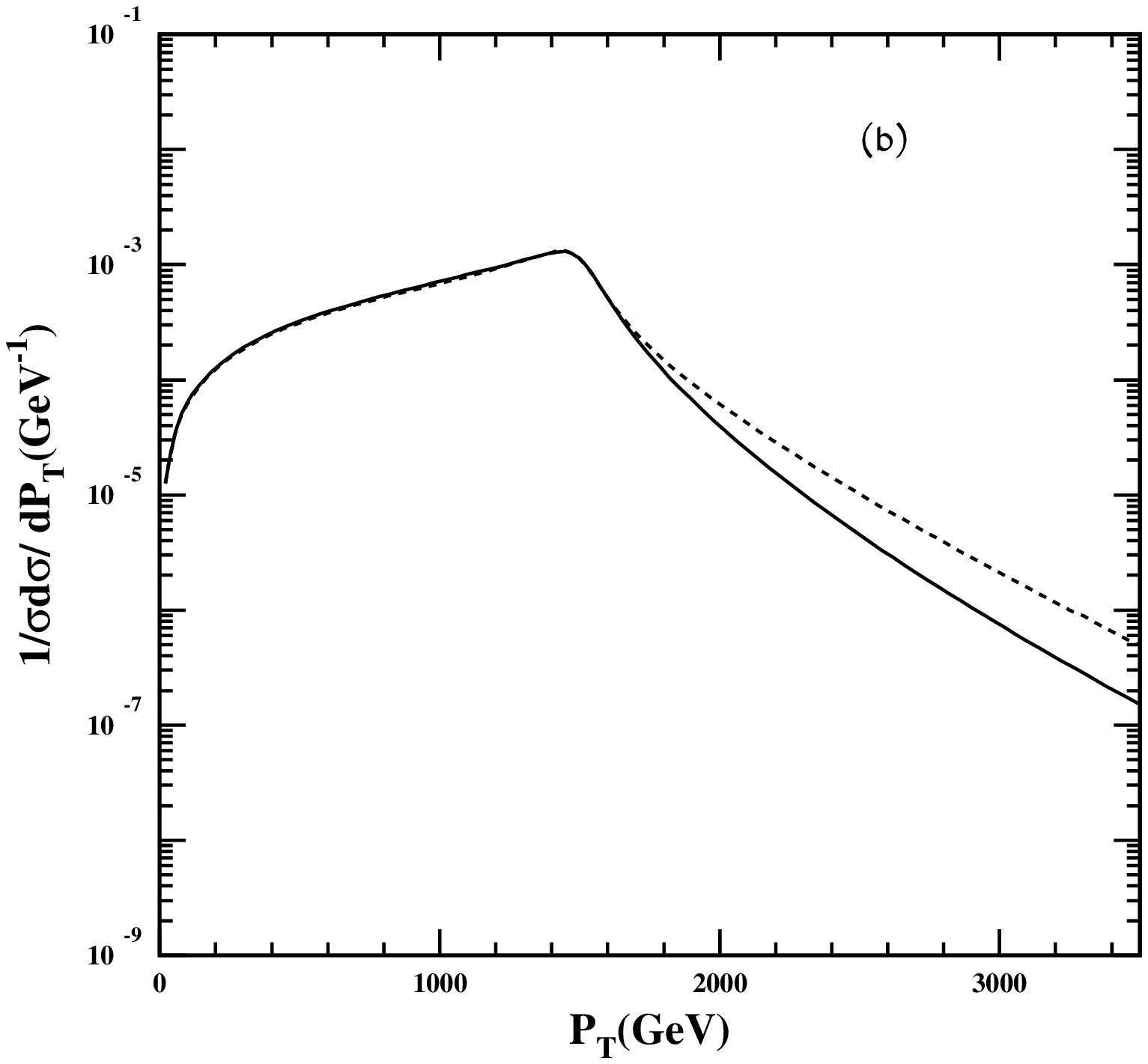}
\includegraphics[width=0.30\textwidth]{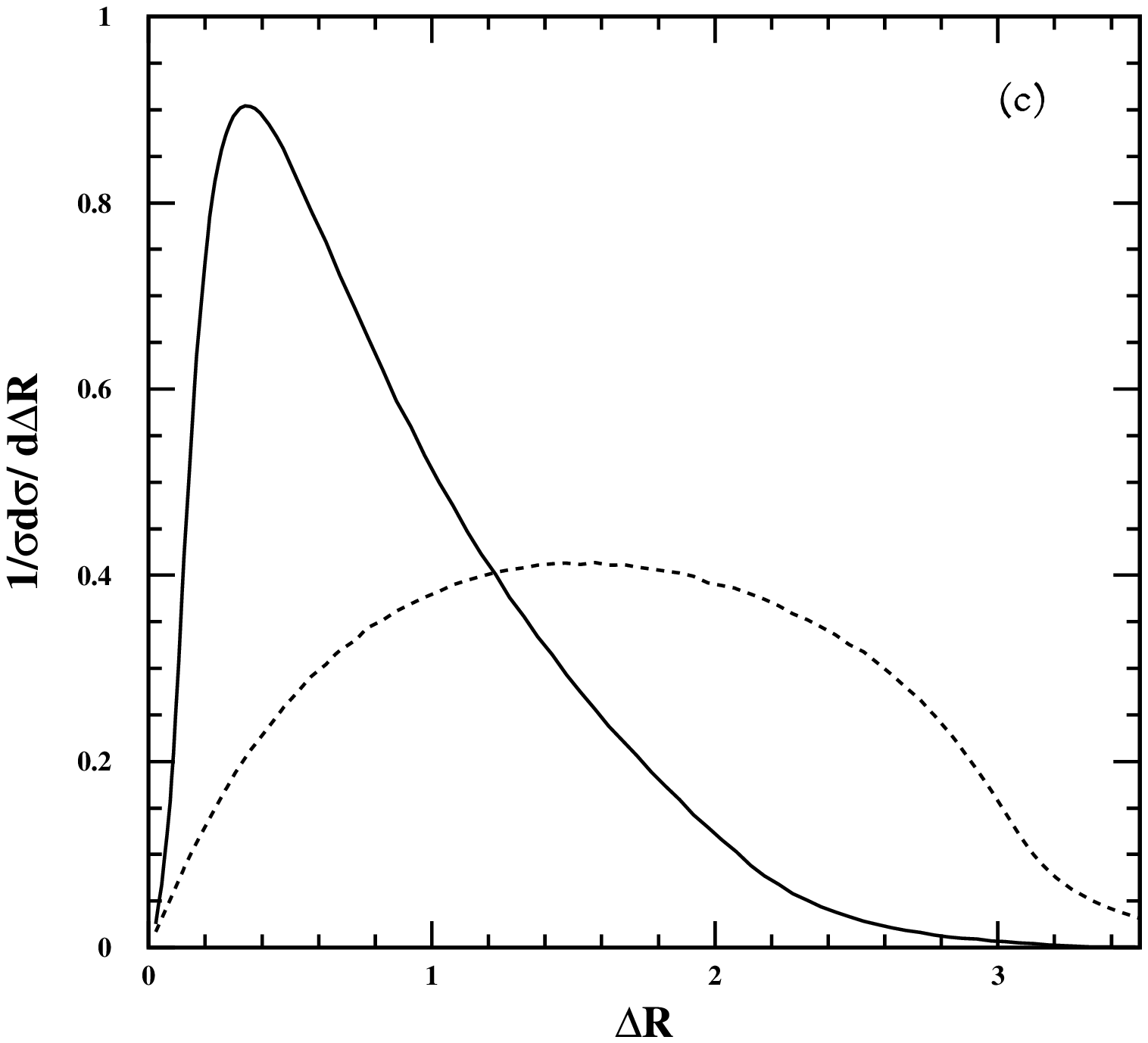}
\caption{(a) The normalized differential distributions with the transverse momentum of the jets ($j_1,j_2,j_3$)
with $p_{Tj_1}>p_{Tj_2}>p_{Tj_3}$ for $M=3$ TeV in the process of $pp\to tW_L^{\prime-}\to 3{\rm jets}+l^{-}+\met$ at $\sqrt{s}=33$ TeV.
(b) The same as (a) but for the charged lepton (solid line) and transverse missing energy $\met$ (dashed line).
(c) The minimal angular separation distributions between jets (solid line) and that between jets and the charged lepton (dashed line).}
\label{fig:pt-lv}
\end{figure}

The distributions of $\sigma^{-1}d\sigma/dP_T$ with the transverse momentum of the three jets and lepton are shown in Fig.~\ref{fig:pt-lv} (a)
and (b). There is an obvious jacobian peak at $P_T=M/2$ in the charged lepton transverse momentum distribution.
In Fig.~\ref{fig:pt-lv} (c), we display the corresponding distributions for $\Delta R$.

We employ the basic acceptance cuts as

\begin{itemize}
\item Cut $C_{1}$:
\begin{flalign}
&{\rm For~~ 14~~ TeV}\left\{\begin{array}{ll}
& p_{lT}>200~{\rm GeV},~~~~\met>200~{\rm GeV} , ~~~~p_{jT}>20~{\rm GeV},\nonumber \\
&|\eta_{l}|<2.5,~~~~|\eta_{j}|<2.5,~~~~\Delta R_{jj(lj)}>0.4.\nonumber\end{array} \right.&\\
&{\rm For~~ 33~~ TeV}\left\{\begin{array}{ll}
& p_{lT}>350~{\rm GeV},~~~~\met>350~{\rm GeV} , ~~~~p_{jT}>20~{\rm GeV},\nonumber \\
&|\eta_{l}|<2.5,~~~~|\eta_{j}|<2.5,~~~~\Delta R_{jj(lj)}>0.4.\nonumber\end{array} \right.&
\end{flalign}
\end{itemize}

And we require the top quark mass be reconstructed by all of the three jets as well as $W$ boson mass reconstructed by two of the jets as the further cut
\begin{itemize}
\item Cut $C_{2}$:
$~~|M_{j_{a}j_{b}}-m_{W}|\leq10~{\rm GeV},~~~~|M_{j_{a}j_{b}j_{c}}-m_{t}|\leq30~{\rm GeV},$
\end{itemize}
where
\begin{equation}
M_{jjj}^{2}=(\sum_{j=1}^{3}{p}_{j})^2.
\end{equation}
In addition, the jet which is not used to reconstruct $W$ boson is tagged as a b-jet to suppress the background processes.

\begin{figure}
\centering \includegraphics[width=0.40\textwidth]{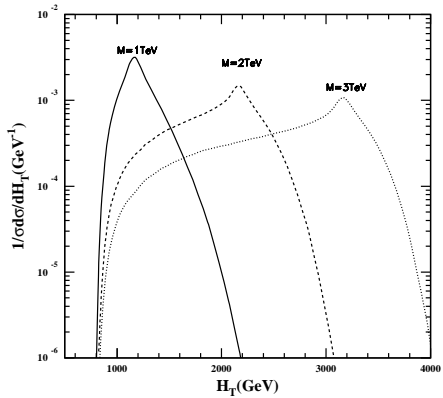}
\caption{The distributions of $1/\sigma d\sigma/dH_T$ with respect to
$H_T$ for $M=1,~2,~3$ TeV for the process of
 $pp\to tW_L^{\prime-}\to l^{-}+3jets+\met$ after cut $C_{2}$ at the LHC for 33 TeV.}
\label{fig:sumpt}
\end{figure}

\begin{figure}
\centering
\includegraphics[width=0.4\textwidth]{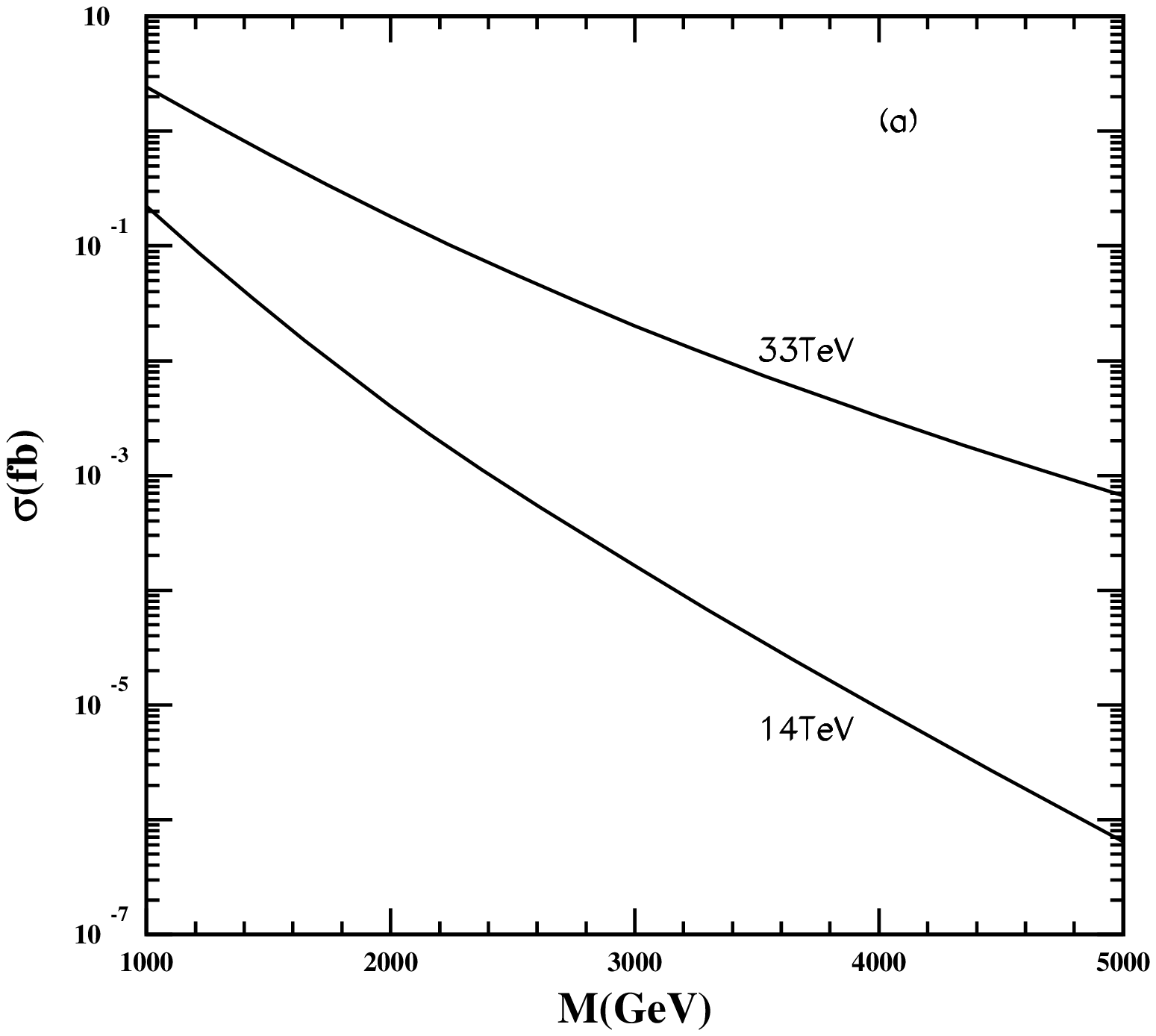}
\includegraphics[width=0.4\textwidth]{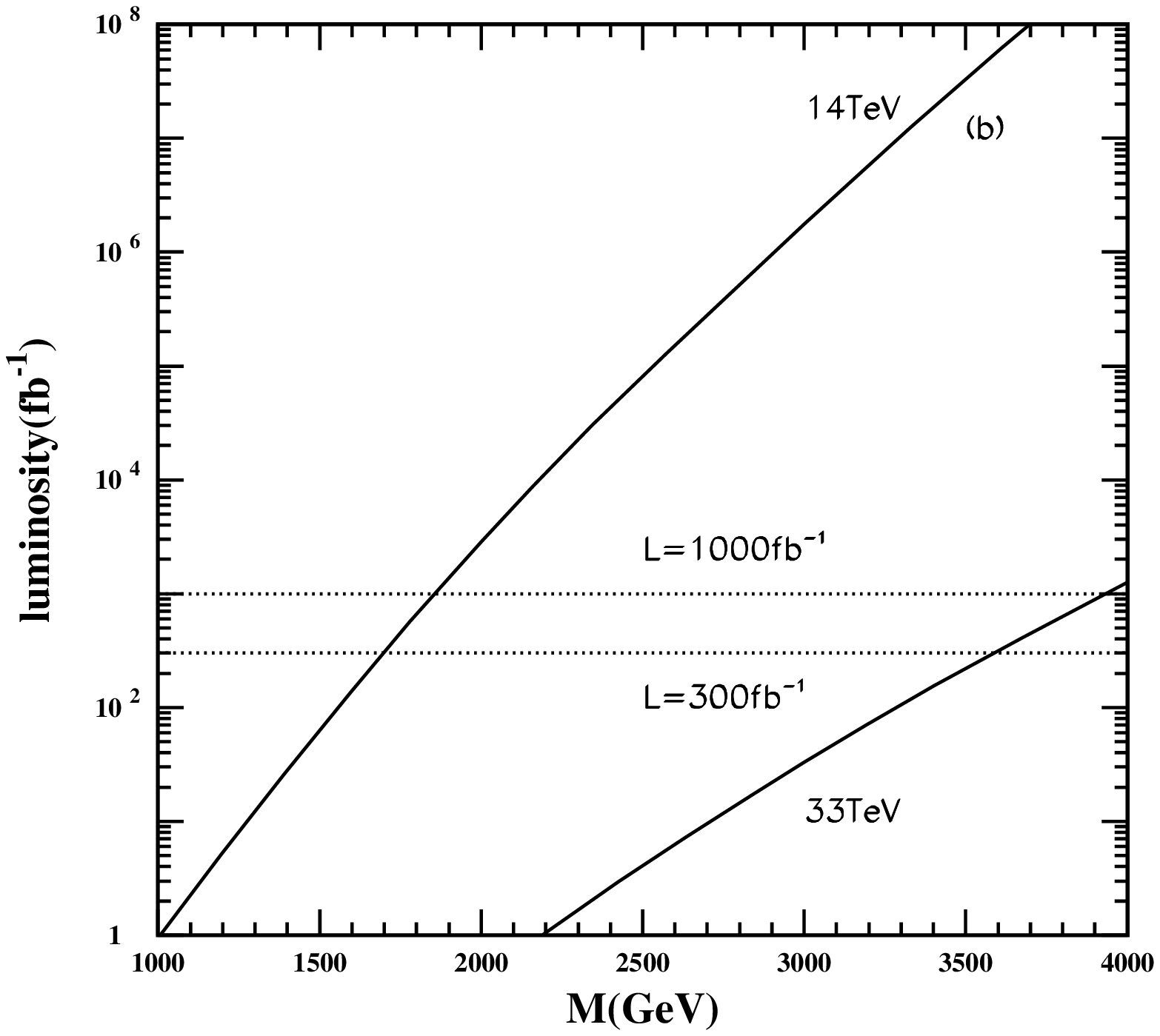}
\caption{(a) The total cross section  as a function of
the charged gauge bosons mass $M$ at $\sqrt{s}=14$ and $33$ TeV
for the process of $pp\to tW^{\prime-}_L\to  3{\rm jets}+l^{-}+\met$.
(b) The integral luminosity needed at the LHC for $\sqrt{s}=14$ and $33$ TeV
at $S/\sqrt{B}=3\sigma$ sensitivity. The two straight dotted lines present the luminosity of $300$ and $1000~fb^{-1}$.}
\label{fig:lum-lv}
\end{figure}

The differential distributions $\sigma ^{-1}d\sigma/dH_T$ for process (\ref{lvchannel}) are
shown in Fig.~\ref{fig:sumpt}, where
\begin{equation}
H_{T}=\sum_{j=1}^{3}{p}_{jT}+{p}_{lT}+\met.
\end{equation}
One can see that a peak appears around the $W^\prime$ mass which is a significant excess comparing to the SM backgrounds.
In Fig.~\ref{fig:lum-lv}, we present the total cross section for the pure left-handed $W^\prime$ with
and without cuts at the LHC with $14$ TeV and $33$ TeV. The cross section reaches 0.02 $fb$ for $M=3$ TeV with $\sqrt{s}=33$ TeV.
Corresponding to the process (\ref{lvchannel}) with final states $3{\rm jets}+l^{-}+\met$, the dominant backgrounds from SM are $tW$, $Wjjj$, $Wjjj$ and $WZj$, which are simulated by MadEvent.
Supposing the integral luminosity to be $300 fb^{-1}$ at $14$ TeV, a $W^\prime _L$ with mass up to 1.7 TeV can be found at $3\sigma$ significance, which can be enlarged to 4 TeV on the condition of integral luminosity of $1000 fb^{-1}$ at the LHC with $\sqrt{s}=33$ TeV.

In a more appealing seesaw version \cite{Minkowski:1977sc,Mohapatra:1979ia} of the theory, as in the papers below, $W^\prime_R$ decays into the heavy right-handed neutrino with the spectacular signatures of lepton number violation ~\cite{Keung:1983uu}. In this case, there have been dedicated searches and theoretical discussions \cite{Ferrari:2000sp,Nemevsek:2011hz,Han:2012vk}. Perhaps the $W^\prime$ can be first detected at the LHC associated with the heavy right-handed neutrino.

\section{Summary}
The observation of a new charged vector boson $W^\prime$ is an unambiguous signal for new physics beyond the SM. As a result, it is important to search for
$W^\prime$ production signal and its related phenomena via different production and decay channels at the LHC.
In this paper, we focus on investigating the collider signature for
the $tW^\prime$ associated production with $W^\prime \to tb$, $W^\prime \to q \bar {q}'$ and $W^\prime \to l \nu$ respectively at the LHC.
It is found that with the most common coupling parameters and present mass constraints for $W^\prime$, due to the limited cross section, it is difficult
to observe $W^\prime$ production signal via $tW^\prime$ associated production at 14 TeV. However, if the center of mass energy of the LHC is updated to 33 TeV or even higher, a distinct signal for $W^\prime$ production can be observed via $W^\prime \to q \bar {q}'$ and $W^\prime \to l \nu$ after adopting proper kinematic cuts, e.g., large transverse momentum cut to the jets/leptons, etc. Once the $tW^\prime$ production is observed at the LHC, it will become important to study the interactions between $W^\prime$ and fermions. For this aim, as an example, we analyze the angular distribution of the charged lepton for
$pp \to tW^{\prime -} \to bW^+ W^{\prime -} \to bl^+ + jj + \met $ process and the related forward-backward asymmetry induced by top quark spin. Our results show that the charged lepton angular distribution is related to the chiral couplings of $W^\prime$ to fermions and the forward-backward asymmetry depending on
this angular distribution. These observables can be used to distinguish $W^\prime_L$ from $W^\prime_R$. If the LHC can served as a discovery machine for the new charged gauge boson $W^\prime$, our work will be useful to search for its production signal and to explore its properties.

\section*{Acknowledgments}
We would like to thank Profs. Shi-Yuan Li and Shou-Shan Bao for their helpful discussions and comments. This work is  supported in part by National Science Foundation of China (NSFC) under grant
Nos.11305075, 11325525, 11275114, 10935012 and 11375200. Li is also supported in part by Natural Science Foundation of Shandong Province under grant No.ZR2013AQ006.



\begin{thebibliography}{37}

\bibitem{Klein:1926tv}
  O.~Klein,
  Z.\ Phys.\  {\bf 37}, 895 (1926),
  [Surveys High Energ.\ Phys.\  {\bf 5} (1986) 241].
\bibitem{ArkaniHamed:1998rs}
  N.~Arkani-Hamed, S.~Dimopoulos and G.~R.~Dvali,
  Phys.\ Lett.\  B {\bf 429}, 263 (1998)
  [arXiv:hep-ph/9803315].
\bibitem{Randall:1999vf}
  L.~Randall and R.~Sundrum,
  Phys.\ Rev.\ Lett.\  {\bf 83}, 4690 (1999); {\bf 83}, 3370 (1999).
\bibitem{ArkaniHamed:2001ca}
  N.~Arkani-Hamed, A.~G.~Cohen and H.~Georgi,
  Phys.\ Rev.\ Lett.\  {\bf 86}, 4757 (2001)
  [arXiv:hep-th/0104005].


\bibitem{Pati:1973uk}
  J.~C.~Pati and A.~Salam,
  Phys.\ Rev.\ D {\bf 8}, 1240 (1973).

\bibitem{Georgi:1974sy}
  H.~Georgi and S.~L.~Glashow,
  Phys.\ Rev.\ Lett.\  {\bf 32}, 438 (1974).

\bibitem{Fritzsch:1974nn}
  H.~Fritzsch and P.~Minkowski,
  Annals Phys.\  {\bf 93}, 193 (1975).


\bibitem{Pati:1974yy}
  J.~C.~Pati and A.~Salam,
  Phys.\ Rev.\  D {\bf 10}, 275 (1974),
  [Erratum-ibid.\  D {\bf 11}, 703 (1975)].

\bibitem{Mohapatra:1974hk}
  R.~N.~Mohapatra and J.~C.~Pati,
  Phys.\ Rev.\  D {\bf 11}, 566 (1975).

\bibitem{Mohapatra:1974gc}
  R.~N.~Mohapatra and J.~C.~Pati,
  Phys.\ Rev.\  D {\bf 11}, 2558 (1975).

\bibitem{Senjanovic:1975rk}
  G.~Senjanovic and R.~N.~Mohapatra,
  Phys.\ Rev.\  D {\bf 12}, 1502 (1975).

\bibitem{Mohapatra:1977mj}
  R.~N.~Mohapatra, F.~E.~Paige and D.~P.~Sidhu,
  Phys.\ Rev.\  D {\bf 17}, 2462 (1978).



\bibitem{Aad:2012dm}
  G.~Aad {\it et al.}  [ATLAS Collaboration],
  Eur.\ Phys.\ J.\ C {\bf 72}, 2241 (2012)
  [arXiv:1209.4446 [hep-ex]].


\bibitem{Chatrchyan:2013lga}
  S.~Chatrchyan {\it et al.}  [CMS Collaboration],
  Phys.\ Rev.\ D {\bf 87}, 072005 (2013)
  [arXiv:1302.2812 [hep-ex]].

\bibitem{Aad:2012ej}
  G.~Aad {\it et al.}  [ATLAS Collaboration],
  Phys.\ Rev.\ Lett.\  {\bf 109}, 081801 (2012)
  [arXiv:1205.1016 [hep-ex]].

\bibitem{Chatrchyan:2012gqa}
  S.~Chatrchyan {\it et al.}  [CMS Collaboration],
  Phys.\ Lett.\ B {\bf 718}, 1229 (2013)
  [arXiv:1208.0956 [hep-ex]].
\bibitem{atlas8tev}
  https://atlas.web.cern.ch/Atlas/GROUPS/PHYSICS/CONFNOTES/ATLAS-CONF-2013-050/



\bibitem{Berger:2011xk}
  E.~L.~Berger, Q.~-H.~Cao, J.~-H.~Yu and C.~-P.~Yuan,
  Phys.\ Rev.\ D {\bf 84}, 095026 (2011)
  [arXiv:1108.3613 [hep-ph]].


\bibitem{Gopalakrishna:2010xm}
  S.~Gopalakrishna, T.~Han, I.~Lewis, Z.~-g.~Si and Y.~-F.~Zhou,
  Phys.\ Rev.\ D {\bf 82}, 115020 (2010)
  [arXiv:1008.3508 [hep-ph]].

\bibitem{Bao:2011nh}
  S.~-S.~Bao, H.~-L.~Li, Z.~-G.~Si and Y.~-F.~Zhou,
  Phys.\ Rev.\ D {\bf 83}, 115001 (2011)
  [arXiv:1103.1688 [hep-ph]].

\bibitem{Bao:2011sy}
  S.~-S.~Bao, X.~Gong, H.~-L.~Li, S.~-Y.~Li and Z.~-G.~Si,
  Phys.\ Rev.\ D {\bf 85}, 075005 (2012)
  [arXiv:1112.0086 [hep-ph]].

\bibitem{Aaltonen:2009qu}
  T.~Aaltonen {\it et al.}  [CDF Collaboration],
  Phys.\ Rev.\ Lett.\  {\bf 103}, 041801 (2009)
  [arXiv:0902.3276 [hep-ex]].


\bibitem{Abazov:2007ah}
  V.~M.~Abazov {\it et al.}  [D0 Collaboration],
  Phys.\ Rev.\ Lett.\  {\bf 100}, 031804 (2008)
  [arXiv:0710.2966 [hep-ex]].

\bibitem{Zhang:2007da}
  Y.~Zhang, H.~An, X.~Ji and R.~N.~Mohapatra,
  Nucl.\ Phys.\ B {\bf 802}, 247 (2008)
  [arXiv:0712.4218 [hep-ph]].

\bibitem{Beall:1981ze}
  G.~Beall, M.~Bander and A.~Soni,
  Phys.\ Rev.\ Lett.\  {\bf 48}, 848 (1982).

\bibitem{Maiezza:2010ic}
  A.~Maiezza, M.~Nemevsek, F.~Nesti and G.~Senjanovic,
  Phys.\ Rev.\ D {\bf 82}, 055022 (2010)
  [arXiv:1005.5160 [hep-ph]].


\bibitem{Chatrchyan:2011ns}
  S.~Chatrchyan {\it et al.}  [CMS Collaboration],
  Phys.\ Lett.\ B {\bf 704}, 123 (2011)
  [arXiv:1107.4771 [hep-ex]].


\bibitem{Chatrchyan:2012su}
  S.~Chatrchyan {\it et al.}  [CMS Collaboration],
  Phys.\ Lett.\ B {\bf 717}, 351 (2012)
  [arXiv:1206.3921 [hep-ex]].


\bibitem{Pumplin:2002vw}
J.~Pumplin, D.~R.~Stump, J.~Huston,H.~L.~Lai, P.~M.~Nadolsky and W.~K.~Tung,
 JHEP \textbf{0207}, 012 (2002) {[}arXiv:hep-ph/0201195{]}. 



\bibitem{atlas0901}
  G.~Aad {\it et al.}  [ATLAS Collaboration],
  arXiv:0901.0512 [hep-ex].

\bibitem{MadEvent} J.~Alwall, M.~Herquet, F.~Maltoni, O.~Mattelaer
and T.~Stelzer, 
 JHEP \textbf{1106}, 128 (2011) {[}arXiv:1106.0522 {[}hep-ph{]}{]}.

\bibitem{Minkowski:1977sc}
  P.~Minkowski,
  Phys.\ Lett.\ B {\bf 67}, 421 (1977).


\bibitem{Mohapatra:1979ia}
  R.~N.~Mohapatra and G.~Senjanovic,
  Phys.\ Rev.\ Lett.\  {\bf 44}, 912 (1980).

\bibitem{Keung:1983uu}
  W.~-Y.~Keung and G.~Senjanovic,
  Phys.\ Rev.\ Lett.\  {\bf 50}, 1427 (1983).

\bibitem{Ferrari:2000sp}
  A.~Ferrari, J.~Collot, M-L.~Andrieux, B.~Belhorma, P.~de Saintignon, J-Y.~Hostachy, P.~.Martin and M.~Wielers,
  Phys.\ Rev.\ D {\bf 62}, 013001 (2000).

\bibitem{Nemevsek:2011hz}
  M.~Nemevsek, F.~Nesti, G.~Senjanovic and Y.~Zhang,
  Phys.\ Rev.\ D {\bf 83}, 115014 (2011)
  [arXiv:1103.1627 [hep-ph]].

\bibitem{Han:2012vk}
  T.~Han, I.~Lewis, R.~Ruiz and Z.~-g.~Si,
  Phys.\ Rev.\ D {\bf 87}, 035011 (2013)
  [Erratum-ibid.\ D {\bf 87}, no. 3, 039906 (2013)]
  [arXiv:1211.6447 [hep-ph]].


\end{thebibliography}
\end{document}